\documentclass[12pt]{article}
\pdfoutput=1

\usepackage{amsmath,amssymb,amsfonts}
\usepackage{psfrag}
\usepackage{color}
\definecolor{darkblue}{rgb}{0.1,0.1,.7}
\usepackage[colorlinks, linkcolor=darkblue, citecolor=darkblue, urlcolor=darkblue, linktocpage]{hyperref} 
\usepackage[square, comma, sort&compress,numbers]{natbib}
\usepackage[]{amsmath}
\usepackage[]{graphicx}
\usepackage[]{latexsym}
\usepackage{geometry}
\newcommand\SDPB{\texttt{SDPB}}
\usepackage{amscd}
\usepackage[all,cmtip]{xy}
\usepackage{mathrsfs}
\usepackage[margin=10pt,font=small,labelfont=bf]{caption}
\usepackage{dsdshorthand}
\usepackage{simplewick}
\usepackage{bbold}
\usepackage{changepage}
\usepackage{slashed}
\usepackage{pstool}
\usepackage{setspace}
\numberwithin{equation}{section}

\renewcommand{\be}{\begin{eqnarray}}
\renewcommand{\ee}{\end{eqnarray}}
\newcommand{\bea}{\begin{eqnarray}}
\newcommand{\eea}{\end{eqnarray}}

\newcommand{\eps}{\epsilon}
\def\beq{\begin{equation}} 
\def\eeq{\end{equation}} 
\def\<{\langle}
\def\>{\rangle}

\def\nn{\nonumber} 
 
\def\cO {{\cal O}}

\def\cN {{\cal N}}

\newcommand\numax{\kappa}

\newcommand{\es}[2] {\begin{equation} \label{#1} \begin{split} #2 \end{split} \end{equation}}

\newcommand{\abs}[1]{\left\lvert #1 \right\rvert}
\def\tr{\text{tr}}

\begin{document}

\vspace*{-.6in} \thispagestyle{empty}
\begin{flushright}
PUPT-2480
\end{flushright}
\vspace{.2in} {\Large
\begin{center}
{\bf Bootstrapping 3D Fermions\\\vspace{.1in}}
\end{center}
}
\vspace{.2in}
\begin{center}
{\bf 
Luca Iliesiu$^{a}$,
Filip Kos$^{b}$, 
David Poland$^{b}$,\\
Silviu S.~Pufu$^{a}$,
David Simmons-Duffin$^{c}$,
Ran Yacoby$^{a}$
} 
\\
\vspace{.2in} 
$^a$ {\it  Joseph Henry Laboratories, Princeton University, Princeton, NJ 08544}\\
$^b$ {\it  Department of Physics, Yale University, New Haven, CT 06520}\\
$^c$ {\it School of Natural Sciences, Institute for Advanced Study, Princeton, New Jersey 08540}
\end{center}

\vspace{.2in}

\begin{abstract}
We study the conformal bootstrap for a 4-point function of fermions $\<\psi\psi\psi\psi\>$ in 3D. We first introduce an embedding formalism for 3D spinors and compute the conformal blocks appearing in fermion 4-point functions. Using these results, we find general bounds on the dimensions of operators appearing in the $\psi \times \psi$ OPE, and also on the central charge $C_T$. We observe features in our bounds that coincide with scaling dimensions in the Gross-Neveu models at large $N$. We also speculate that other features could coincide with a fermionic CFT containing no relevant scalar operators.
\end{abstract}

\newpage

\begin{spacing}{0.8}
\tableofcontents
\end{spacing}

\newpage

\section{Introduction}
\label{sec:intro}

The conformal bootstrap \cite{Polyakov:1974gs,Ferrara:1973yt,Mack:1975jr}, originally successful in elucidating 2D conformal field theories (CFTs), has recently become a powerful method to constrain the operator algebra of unitary CFTs also in $D>2$ spacetime dimensions. The origin of this development is the observation made in \cite{Rattazzi:2008pe} that the combined constraints of crossing symmetry and unitarity on a 4-point function of scalars can be explored numerically. This method achieved an impressive degree of success, for example, by enabling accurate determinations of the dimensions of low-lying operators in particular CFTs, such as the 3D Ising \cite{ElShowk:2012ht,El-Showk:2014dwa,Kos:2014bka,Simmons-Duffin:2015qma} and critical $O(N)$ vector \cite{Kos:2013tga,Kos:2015mba} models. While the original ideas of \cite{Rattazzi:2008pe} have been developed in myriad ways in subsequent works, the basic objects of study have always been 4-point functions of scalar operators. The goal of the present paper is to extend the bootstrap toolbox to study 4-point functions of fermionic operators in 3D CFTs.

There are many motivations for bootstrapping fermionic correlators. Fermionic operators exist in many interesting CFTs, though they do not appear in the operator product expansion (OPE) of scalar operators. Therefore, in order to access this sector of the operator algebra, one must study correlators of fermionic operators.  More generally, it is of great interest to apply the conformal bootstrap to 4-point functions of operators with non-zero spin. For example, studying the 4-point function of a global symmetry current would lead to universal bounds on all CFTs that admit the corresponding symmetry, without making any additional assumptions on their operator content. Similarly, bootstrapping the 4-point function of the stress-tensor would allow for the most general constraints, since the stress-tensor exists in any local CFT by definition.\footnote{In some supersymmetric theories correlators of symmetry currents are related to correlators of scalars. The numerical bootstrap has been applied to such cases in \cite{Beem:2013qxa,Berkooz:2014yda,Chester:2014fya,Chester:2014mea,Beem:2014zpa,Beem:2015aoa}.} Our numerical study of fermion correlators can be seen as a small step towards implementing the bootstrap for operators with spin, as in the examples discussed above.

In implementing the bootstrap for correlators of fermionic operators, or of operators with spin more generally, one faces two difficulties. The first is that the explicit form of the conformal block decomposition for higher-spin correlators is not known in general for $D>2$ dimensional CFTs. The exception occurs in $D=3$, where the conformal blocks for external operators with integer spin can be determined by acting with certain differential operators on the scalar blocks \cite{Costa:2011dw}.  As we will see, a similar strategy applies for operators with half-integer spin. The second difficulty has to do with the proliferation of conformal invariants that can appear in higher-spin correlators. In particular, the number of conformal invariants grows with the spin of the insertions (see, for instance, \cite{Giombi:2011rz,Costa:2011mg}).  As we will explain, a 4-point function of primary operators of spin-$1/2$ generally depends on $16$ independent conformal invariants. Imposing parity symmetry reduces the number of invariants to $8$, and this number can be reduced further to $5$ if we assume the fermions are identical. In our numerical analysis, we will focus for simplicity on this latter case and derive the conformal block decomposition of a 4-point function of identical fermionic operators in a 3D CFT. 

With the conformal blocks in hand, we then embark on a systematic study of CFTs with a small number of relevant operators.  By imposing gaps in the low-lying spectrum of parity-even and/or parity-odd scalar operators, we use the logic originally introduced in \cite{Rattazzi:2008pe} to derive constraints on the dimensions of the first few scalar operators and on that of the fermionic operator whose 4-point function we study.  We also find a lower bound on the coefficient $C_T$ that appears in the two-point function of the canonically-normalized stress tensor.

From these studies we find two exciting results. Firstly, the general bound on the dimension of the leading parity-odd scalar $\Delta_{\sigma}$ possesses a severe discontinuity at a fermion dimension of $\Delta_{\psi} \sim 1.27$. This coincides with a kink in the general bound on the leading parity-even scalar dimension $\Delta_{\epsilon}$.  Based on these features we conjecture the existence of a fermionic 3D CFT containing no relevant scalar operators. While we do not know of a Lagrangian that would give rise to such a theory, this conjectured ``dead-end" CFT would furnish a concrete example of self-organized criticality~\cite{Bak:1987xua,Bak:1988zz} in 3D.

The second result is that, when we allow a second relevant parity-odd scalar in the spectrum with dimension $\Delta_{\sigma'} = 2 + \delta$ for small values of $\delta$, the resulting allowed region for $(\Delta_{\psi}, \Delta_{\sigma})$ possesses a sharp kink that appears to precisely coincide with the dimensions in the $O(N)$ Gross-Neveu models at large $N$. This is natural because in the large-$N$ limit one has the expansion $\Delta_{\sigma'} = 2 + 32/(3\pi^2N) + \ldots$. By tracking this feature at larger values of the gap we reveal information about the small-$N$ Gross-Neveu models, including the $N=1$ theory which is expected to have $\cN=1$ supersymmetry. In addition to this sequence of kinks, we observe the emergence of a second discontinuity in the allowed region at $(\De_\psi,\De_\s)\approx(1.078,0.565)$, which we conjecture could also coincide with a 3D CFT with fermionic operators and a large scalar gap. We believe that fully isolating these theories will require implementing systems of mixed correlators containing fermions and scalars, but based on the results of this study the prospects for learning more about these theories using the conformal bootstrap looks very promising.

The rest of this paper is organized as follows.  In Section~\ref{EMBEDDING} we start by reviewing the embedding space formalism for operators with spin and, using this formalism, derive the conformal block decomposition of a 4-point function of identical fermions.  In Section~\ref{sec:bootstrap} we set up the crossing equations and outline the strategy we will follow in our numerical study.  Next, in Section~\ref{sec:results}, we present numerical results for bounds on dimensions of scalar operators in theories with fermions that satisfy various assumptions, and also present our lower bound on $C_T$.  We end in Section~\ref{sec:discussion} with a discussion of our results.

\section{Embedding Formalism for 3D Spinors}
\label{EMBEDDING}

In order to set up the 3D fermion bootstrap, we need an efficient formalism for keeping track of the tensor structures appearing in correlators of fermionic operators.  We also need to calculate the conformal blocks appearing in the expansion of fermion 4-point functions.  Our approach will be to use an embedding formalism where the spinorial 3D conformal group $\Sp(4,\bR)$ is linearly realized. Similar CFT embedding methods have been developed in~\cite{Dirac:1936fq,Mack:1969rr,Boulware:1970ty,Ferrara:1973eg,Weinberg:2010fx,Giombi:2011rz,Costa:2011mg,Costa:2011dw,SimmonsDuffin:2012uy,Costa:2014rya,Elkhidir:2014woa,Echeverri:2015rwa} and various supersymmetric extensions have been developed recently in~\cite{Siegel:2010yd, Goldberger:2011yp, Siegel:2012di, Maio:2012tx, Kuzenko:2012tb, Goldberger:2012xb, Khandker:2012pa, Fitzpatrick:2014oza} and references therein. Details of our group theory conventions are given in Appendix~\ref{app:grouptheory}.

\label{sec:embedding}

We label the 3D coordinates as $x^{\mu}$, with $\mu = 0, 1, 2$, and use the Minkowski metric in mostly plus signature $\eta^{\mu\nu} = \eta_{\mu\nu}= \textrm{diag}(-1,1,1)$. The coordinates $x^\mu$ transform non-linearly under special conformal transformations. It is therefore convenient to introduce a different set of coordinates that transform linearly under the action of conformal transformations. Since the 3D conformal group is isomorphic to $\SO(3,2)$, we can relate conformal transformations to Lorentz transformations in a 5D spacetime with metric $\eta^{AB} = \eta_{AB} = \textrm{diag}(-1,1,1,1,-1)$, where the indices $A, B$ run from $0$ to $4$.  The exact relation between the generators of conformal transformations and the $\SO(3,2)$ generators $J^{AB}$ can be taken to be
\be
D=-J^{34}\,, \quad P^\mu = J^{3\mu} + J^{4\mu}\,, \quad K^\mu =- J^{3\mu}+J^{4\mu}\,, \quad M^{\mu\nu} = J^{\mu\nu}\,.
\ee

Let us denote 5D coordinates that transform linearly under $\SO(3, 2)$ by capital letters, $X^A$. The way to embed the 3D coordinates $x^\mu$ in 5D space is through the projective null cone, which is defined as the space of all points $X^A$ that satisfy the condition $X \cdot X =0$ and are identified up to a rescaling $X^A \sim \lambda X^A$. It will be convenient to use  lightcone coordinates $X^\pm = X^4 \pm X^3$, and thus write the 5D coordinates from now on as $X = (X^\mu, X^+, X^-)$.  The exact relation between $x^\mu$ and $X^A$ is given by
\be
\label{eq:projectivelightcone}
x^\mu = \frac{X^\mu}{X^+}\,, \qquad X=  X^+(  x^\mu ,1 ,  x^2 )\,,
\ee
where $x^2 \equiv \eta_{\mu\nu} x^\mu x^\nu$.  Note that the parameterization \eqref{eq:projectivelightcone} obeys $X \cdot X = 0$.

\subsection{Embedding of Scalar Fields}

To find the embedding of fields in 5D spacetime we follow the approach of \cite{Weinberg:2010fx}. Consider first a real scalar primary field $\phi(x)$. Its transformations under the conformal group are
 \es{phiConf}{
  i [M^{\mu\nu}, \phi(x)] &= (x^\nu \partial^\mu - x^\mu \partial^\nu) \phi(x) \,, \\
  i [P^\mu, \phi(x)] &= - \partial^\mu \phi(x) \,, \\
  i [K^\mu, \phi(x)] &= \left( 2 x^\mu x^\nu \partial_\nu - x^2 \partial^\mu + 2 \Delta_\phi x^\mu  \right) \phi(x) \,, \\
  i [D, \phi(x) ] &= \left( x^\mu \partial_\mu + \Delta_\phi \right) \phi(x) \,,
 }
where $\Delta_\phi$ is the dimension of $\phi$. We can relate $\phi(x)$ to a scalar field $\Phi(X)$ defined on the lightcone in 5D as:
\be
\label{eq:embeddingphi}
\Phi (X) = \frac{1}{(X^+)^{\Delta_\phi}} \phi(x)\,,
\ee
where $x$ is related to $X$ through (\ref{eq:projectivelightcone}).  Explicit calculation then shows that $\Phi(X)$ is a 5D Lorentz scalar, i.e.~that it transforms under 5D Lorentz transformations as
\be
i[J^{AB},\Phi(X)] = \left( X^B\frac{\partial}{\partial X_A} -  X^A\frac{\partial}{\partial X_B}   \right) \Phi(X)\,,
\ee
if and only if $\phi(x)$ is a primary scalar field in 3D with dimension $\Delta_\phi$. 

Note that the 5D field $\Phi(X)$ defined in (\ref{eq:embeddingphi}) is a homogeneous function of $X$ of degree $-\Delta_\phi$. This property together with 5D Lorentz invariance restricts the form of correlation functions in embedding space.  For instance, the two-point function takes the form
  \es{TwoPtScalar}{
  \langle \Phi(X_1) \Phi(X_2) \rangle = \frac{c_\phi}{X_{12}^{\Delta_\phi}} \,, \qquad X_{ij} \equiv -2 X_i \cdot X_j \,.
  }
Using \eqref{eq:embeddingphi}, we can read off
 \be
  \langle \phi(x_1) \phi(x_2) \rangle= \frac{c_\phi}{x_{12}^{2 \Delta_\phi}} \,, \qquad x_{ij}^\mu \equiv x_i^\mu - x_j^\mu  \,,
 \ee
as expected.  In a unitary theory, we must have $c_\phi>0$.  We will conventionally take $c_\phi=1$ for scalar operators in this work.

\subsection{Embedding of Spinor Fields}

The above procedure can be applied to primary spinor fields as well. The main difference is that such a field $\psi^{\alpha}(x)$ transforms in a spinor representation of the double cover of $\SO(2,1)$, which is isomorphic to $\Sp(2,\bR)$.  Under the full conformal group, it transforms as
 \es{psiConf}{
  i [M^{\mu\nu}, \psi^\alpha(x)] &= (x^\nu \partial^\mu - x^\mu \partial^\nu) \psi^\alpha(x) - i(\mathcal{M}^{\mu\nu})^\alpha{}_\beta \psi^\beta(x) \,, \\
  i [P^\mu, \psi^\alpha(x)] &= - \partial^\mu \psi^\alpha(x) \,, \\
  i [K^\mu, \psi^\alpha(x)] &= \left( 2 x^\mu x^\nu \partial_\nu - x^2 \partial^\mu + 2 \Delta_\psi x^\mu  \right) \psi^\alpha(x) +2 i x_\nu ({\cal M}^{\nu\mu})^\alpha{}_\beta \psi^\beta(x) \,, \\
  i [D, \psi^\alpha(x) ] &= \left( x^\mu \partial_\mu + \Delta_\psi \right) \psi^\alpha(x)  \,,
 }
where $\mathcal{M}^{\mu\nu} = -\frac{i}{4}[\gamma^\mu,\gamma^\nu]$.  Here, upper (lower) indices $\alpha,\beta, \ldots,$ represent fundamental (anti-fundamental) $\Sp(2, \bR)$ indices that are raised and lowered with the symplectic form $\Omega_{\alpha\beta} = \Omega^{\alpha\beta}$---see Appendix~\ref{app:grouptheory} for our conventions.  In 3D, the smallest spinor representation is a 2-component Majorana spinor, which is what we will focus on.   If we take the 3D $\gamma$ matrices to be real, as we do in Appendix~\ref{app:grouptheory}, a Majorana spinor has real components, $\psi^\alpha(x)^* = \psi^\alpha(x)$.

In order to efficiently keep track of the 3D spinor indices, it is convenient to introduce a set of auxiliary commuting variables $s_\alpha$ and consider the product
 \be
  \psi(x, s) \equiv s_\alpha \psi^\alpha(x) \,. \label{psixsDef}
 \ee
The quantity $\psi(x, s)$ contains the same information as the spinor fields $\psi^\alpha(x)$, because the latter can be recovered through $\psi^\alpha(x) = \frac{\partial}{\partial s_\alpha} \psi(x, s)$.

Going to the embedding space, we use the double cover of $\SO(3,2)$, which is isomorphic to $\Sp(4,\bR)$ with generators $\mathcal{M}^{AB} = -\frac{i}{4}[\Gamma^A,\Gamma^B]$.   For every 3D spinor field $\psi^\a(x)$, we would like to define a 5D spinor field $\Psi^I(X)$  on the lightcone \eqref{eq:projectivelightcone}, where upper (lower) indices $I, J$, etc.~denote fundamental (anti-fundamental) $\Sp(4, \bR)$ indices that are raised and lowered with the symplectic form $\Omega_{IJ} = \Omega^{IJ}$.  As in 3D, we can also introduce polarization variables $S_I$ that help us efficiently keep track of the $\Sp(4, \bR)$ indices,
 \be
  \Psi(X, S) \equiv S_I \Psi^I(X) \,. \label{PSIXSDef}
 \ee
If we wish, we are free to treat $S_I$ as a spurionic 5D field that transforms in the anti-fundamental of $\Sp(4, \bR)$, and similarly to treat $s_\a$ as a position-independent spurionic field in 3D that transforms in the anti-fundamental of $\Sp(2, \bR)$ and is invariant under dilatations (in other words, $s_\a$ is a primary spinor field of vanishing dimension).  If we do so, then $\Psi(X, S)$ and $\psi(x, s)$ become 5D and 3D Lorentz scalars, respectively.  Just as in \eqref{eq:embeddingphi}, one can check that the relation 
 \be
  \Psi(X, S) = \frac{1}{(X^+)^{\Delta_\psi}} \psi(x, s)  \label{PsiXSRelation}
 \ee
implies that $\Psi(X, S)$ is a Lorentz scalar in 5D if and only if $\psi(x, s)$ is a primary field in 3D with dimension $\Delta_\psi$.  Since we assumed that $s_\alpha$ transforms as a dimension-zero primary field, then we have that $\Psi(X, S)$ is a 5D scalar if and only if $\psi^\alpha$ is an $\Sp(2, \bR)$ spinor primary field of dimension $\Delta_\psi$.

To finish the identification between the 3D and 5D spinor fields, we can take\footnote{Equivalently, we could have written $\tilde \Psi(X, \tilde S) = \tilde S_I \tilde \Psi^I(X) = \psi(x, s) / (X^+)^{\Delta_\psi}$ and identified
$$ \tilde\Psi^I(X) = \frac{1}{(X^+)^{\Delta_\psi - 1/2}} \begin{pmatrix} -x^\a{}_\b \psi^\b(x)  \\ \psi_\a (x)\end{pmatrix}\,,
$$
such that $\tilde{\Psi}^I(X)$ is a 5D spinor if and only if $\psi^\alpha(x)$ is a primary spinor field of dimension $\Delta_\psi$.  Then $\tilde \Psi(X)$ would satisfy the transversality condition ${X^I}_J \tilde\Psi^J(X) = 0$ and the polarization $\tilde S_I$ would only be defined modulo shifts $\tilde S_I \to \tilde S_I + T_J X^J{}_I$.  One can relate this description to the one presented in the main text by taking $S_I = \tilde S_J X^J{}_I$ and $\tilde \Psi^I(X) = X^I{}_J \Psi^J(X)$.

With this in mind, we can relate our formalism to that of \cite{Costa:2011mg} for traceless symmetric tensor fields.  They define a vector $Z^A$ that satisfies transverseness $Z\.X=0$ and is defined up to gauge redundancy $Z\to Z+\lambda X$.  Such a vector can be obtained as $Z^A=\tl S\Gamma^A \Omega S$.  Note that the gauge redundancy of $\tl S$ gives $Z^A \to Z^A + T X\Gamma^A \Omega S=Z^A + X^A (T\Omega S)$, which is the correct gauge redundancy for $Z$.  Similarly $Z\.X=\tl S X \Omega S=0$ since $S$ is transverse.
} 
  \es{STos}{
  S_I = \sqrt{X^+} \begin{pmatrix}
   s_\alpha \\
   -x^{\alpha}{}_{\beta} s^{\beta} 
  \end{pmatrix} \,,\qquad  x^{\alpha}{}_{\beta}\equiv x^{\mu}(\gamma_{\mu})^{\alpha}{}_{\beta} \,.
  }
Using the conventions of Appendix~\ref{app:grouptheory} for the embedding of $\Sp(2, \bR)$ into $\Sp(4, \bR)$, one can check explicitly that this relation implies that if $S_I$ is an $\Sp(4, \bR)$ anti-fundamental spinor in 5D, then $s_\alpha$ is an $\Sp(2, \bR)$ anti-fundamental spinor primary field in 3D with vanishing dimension, as desired.  Notice that $S_I$ satisfies the transversality condition 
 \be
   S_I X^I{}_J = 0 \,, \qquad  X^I{}_J \equiv X^A (\Gamma_A)^I{}_J \,,  \label{Transversality}
 \ee
which is invariant under $\Sp(4,\bR)$ transformations and is consistent with the lightcone condition $X \cdot X= 0$.  Due to this transversality condition, the 5D spinor field $\Psi^I$ is defined on the lightcone only modulo the shifts $\Psi^I(X) \to \Psi^I(X) + X^I{}_J \Theta^J(X)$, where $\Theta^J(X)$ is an arbitrary spinor on the lightcone.

$\Psi(X, S)$ satisfies the homogeneity property
\begin{align}
\Psi(a X, b S) = a^{-\Delta_{\psi}-1/2}b\Psi(X,S),
\end{align}
where $a$ and $b$ are arbitrary and independent. Homogeneity, the transversality condition \eqref{Transversality}, and 5D Lorentz invariance restrict the form of embedding space correlation functions of $\Psi(X, S)$.  For example, the only consistent expression for the two-point function is
 \be
  \langle \Psi(X_1, S_1) \Psi(X_2, S_2) \rangle = i c_\psi \frac{ \langle S_1 S_2 \rangle}{X_{12}^{\Delta_\psi + \frac 12}} \,, \label{TwoPointFermions}
 \ee 
for some constant $c_\psi$.  Here, we used the notation
\be
\<S_1 X_2 X_3 \dots S_n \> = S_{1I} {{X_2}^I}_J {{X_3}^J}_K \dots \Omega^{LM} S_{nM}\,,
\ee
where ${X^I}_J$ is defined in \eqref{Transversality} and $\Omega_{IJ}=\Omega^{IJ}$ is the $\Sp(4,\bR)$ invariant tensor (see Appendix~\ref{app:grouptheory}).  Using \eqref{PsiXSRelation} and \eqref{STos} in \eqref{TwoPointFermions}, we obtain, as expected
 \be
  \langle \psi^\alpha (x_1) \psi_\beta(x_2) \rangle = i c_\psi \frac{(x_{12})^{\alpha}{}_{\beta} }{x_{12}^{2 \Delta_\psi + 1}} \,.
 \ee
For Majorana fermions, we have $c_\psi \in \bR$, as can easily be seen by using the Majorana condition $\psi^{\alpha*} = \psi^\alpha$ and the fact that complex conjugation interchanges the order of the Grassmann variables.  (Recall that we work in a basis where the gamma matrices are real.)  In this paper, we will take $c_\psi = 1$ for all external operators.

\subsection{Embedding of Fields of Higher Spin}

The above discussion generalizes to fields of higher spin in a natural way. In 3D, a spin-$\ell$ field is a totally symmetric tensor in $2\ell$ spinor indices: $\cO^{\a_1\a_2\dots \a_{2\ell}}(x)$. It corresponds to an embedding field $\cO^{I_1I_2\dots I_{2\ell}}(X)$ that is homogeneous of degree $-(\Delta+\ell)$ in $X$ and also totally symmetric in its 5D spinor indices. In index-free notation, we contract all the indices with an auxiliary spinor $s$ in 3D or with a transverse auxiliary spinor $S$ in 5D:
 \be 
  \cO_\ell(x, s) = s_{\a_1} s_{\a_2} \cdots s_{\a_{2 \ell}} \cO^{\a_1\a_2\dots \a_{2\ell}}(x) \,,  \qquad
   \cO_{\ell}(X,S) = S_{I_1} S_{I_2} \cdots S_{I_{2\ell}}\cO^{I_1I_2\dots I_{2\ell}}(X) \,.
 \ee
 $\cO_{\ell}(X,S)$ is also homogeneous in the variable $S$, with degree $2\ell$.   By the same argument that led to \eqref{PsiXSRelation}, we must have
  \be
   \cO_\ell(X, S) = \frac{1}{(X^+)^{\Delta_{\cal O}}} \cO_\ell(x, s) \,.
  \ee
With the help of \eqref{STos} and
 \es{OComponents}{
  \cO^{\a_1\a_2\dots \a_{2\ell}}(x) = \frac{1}{(2 \ell)!} \frac{\partial^{2 \ell}}{\partial s_{\a_1} \partial s_{\a_2} \ldots \partial s_{\a_{2\ell}}} {\cal O}(x, s) \,,
 }
one can then reconstruct the correlation functions of $\cO^{\a_1\a_2\dots \a_{2\ell}}(x)$ from the corresponding formulas in embedding space.

As an example, the two-point function of $ \cO_\ell(X, S)$ in embedding space is restricted to take the form 
  \be
  \langle \cO_\ell(X_1, S_1) \cO_\ell(X_2, S_2) \rangle = i^{2\ell} c_{\cal O} \frac{ \langle S_1 S_2 \rangle^{2\ell}}{X_{12}^{\Delta_{\cO} + \ell}} \,. \label{TwoPointArbitrarySpin}
 \ee 
Here $c_{\cO}$ is real if $\cO_\ell$ is real.  If $\ell$ is an integer, we also have $c_{\cO} > 0$ in a unitary theory for a real operator ${\cal O}_\ell$.

For future reference, we record that when $\ell$ is an integer, we could have represented the spin-$\ell$ operator in terms of a rank-$\ell$ traceless symmetric tensor of $SO(2,1)$, namely ${\cal O}^{\mu_1\ldots \mu_\ell}$.  This tensor is related to $\cO^{\a_1 \ldots \a_{2\ell}}$ via 
 \es{OSptime}{
  {\cal O}^{\a_1 \ldots \a_{2 \ell}}(x) &= \cO^{\mu_1 \ldots \mu_\ell}(x) \gamma_{\mu_1}^{\a_1 \a_2} \cdots \gamma_{\mu_\ell}^{\a_{2\ell-1} \a_{2\ell}} \,, \\
   \cO^{\mu_1 \ldots \mu_\ell}(x) &=  \frac{(-1)^\ell}{2^\ell}  \gamma^{\mu_1}_{\a_1 \a_2} \ldots  \gamma^{\mu_\ell}_{\a_{2\ell-1} \a_{2\ell}}   {\cal O}^{\a_1 \ldots \a_{2 \ell}}(x) \,.
 }
It is straightforward to show that for spin-1 operators, the 2-point function in \eqref{TwoPointArbitrarySpin} can also be written in the more familiar form
 \es{Spin1CooDrds}{
  \langle \cO^\mu(x_1) \cO^\nu(x_2) \rangle = \frac{c_\cO}{2} \frac{I^{\mu\nu}(x_{12})}{\abs{x_{12}}^{2 \Delta{_\cO}}} \,, 
 }
where $I^{\mu\nu}(x) \equiv \eta^{\mu\nu} - 2 x^\mu x^\nu / x^2$.  The analogous expression for spin-2 operators is
  \es{Spin2Coords}{
  \langle \cO^{\mu\nu}(x_1) \cO^{\rho\sigma}(x_2) \rangle = \frac{c_\cO}{4}\left[ \frac 12 \left( I^{\mu\rho}(x_{12}) I^{\nu\sigma}(x_{12}) + I^{\mu\sigma}(x_{12}) I^{\nu\rho}(x_{12})\right) - \frac 13 \eta^{\mu\nu} \eta^{\rho\sigma}   \right]  \frac{1}{\abs{x_{12}}^{2 \Delta_{\cO}}} \,.
 }
For a conserved current, take ${\cal O} = J$ with $\Delta_{J} = 2$ in \eqref{Spin1CooDrds}, and for the stress tensor take ${\cal O} = T$ with $\Delta_{T} = 3$ in \eqref{Spin2Coords}.

\subsection{Three- and Four-Point Functions}

3-point functions between two scalars and a spin-$\ell$ operator of dimension $\Delta$ are uniquely constrained up to an overall coefficient $\lambda_{\phi_1 \phi_2 \cO}$ to be of the form
 \es{ScalarThreePoint}{
     \<\Phi_1(X_1) \Phi_2(X_2) \cO_{\ell}(X_3,S_3)\> 
        &= \lambda_{\phi_1 \phi_2 \cO} \frac{\<S_3 X_1 X_2 S_3\>^{\ell}}{X_{12}^{\frac{\Delta_1 + \Delta_2 - \Delta + \ell}{2}} 
            X_{23}^{\frac{\Delta_2 - \Delta_1 + \Delta + \ell}{2}} 
            X_{31}^{\frac{\Delta_1 - \Delta_2 + \Delta + \ell}{2}}}  \,. 
  }
This form follows from homogeneity of degree $-\Delta_1$, $-\Delta_2$, $-(\Delta + \ell)$ in $X_1$, $X_2$, and $X_3$, respectively, homogeneity of degree $2 \ell$ in $S_3$, transversality of $S_3$ with respect to $X_3$, and $Sp(4,\bR)$ invariance.  In addition, it is useful to note that $\{X_1, X_2\} = (2 X_1 \cdot X_2) \mathbb{1}_4 $, which, together with $X_k \cdot X_k = 0$, restricts the choice of quantities that can appear in between $\langle S_3 (\cdots) S_3 \rangle$ to what is written in \eqref{ScalarThreePoint}.  Note that 
 \es{ScalarInterchange}{
   \<\Phi_1(X_1) \Phi_2(X_2) \cO_{\ell}(X_3,S_3)\> = (-1)^\ell \<\Phi_1(X_2) \Phi_2(X_1) \cO_{\ell}(X_3,S_3)\> \,,
 }
 so in the case of identical operators $\Phi_1 =\Phi_2$, the 3-point function necessarily vanishes if $\ell$ is odd.  Furthermore, for real scalar operators we have $\<\Phi_1(X_1) \Phi_2(X_2) \cO_{\ell}(X_3,S_3)\>^* = \<\Phi_1(X_1) \Phi_2(X_2) \cO_{\ell}(X_3,S_3)\>$, so the structure constants $\lambda_{\phi_1 \phi_2 \cO}$ appearing in \eqref{ScalarThreePoint} are real.

When we come to 3-point functions containing fermions, we have the new complication that multiple tensor structures can appear.  In general, we have
 \es{FermThreePoint}{
   \<\Psi_1(X_1,S_1) \Psi_2(X_2,S_2) \cO_\ell(X_3,S_3)\> = \frac{\sum_a \lambda_{\psi_1 \psi_2 \cO}^a\, r_a}{X_{12}^{\frac{\Delta_1 + \Delta_2 - \Delta - \ell+1}{2}} X_{23}^{\frac{\Delta_2 - \Delta_1 + \Delta + \ell}{2}} X_{31}^{\frac{\Delta_1 - \Delta_2 + \Delta + \ell}{2}}} \,.
 }
where the index $a$ runs over all possible 3-point structures $r_a$, to be given shortly.  These structures can be divided into those that are even under parity $X_k \rightarrow -X_k$ and those that are odd. A basis for the parity-even structures is given by
\begin{align}
  r_1 &= 
   \frac{\<S_1 S_2\> \<S_3 X_1 X_2 S_3\>^{\ell}}{X_{12}^{\ell} } \,, \\
 r_2 &= \frac{\<S_1 S_3\> \<S_2 S_3\> \<S_3 X_1 X_2 S_3\>^{\ell-1}}{X_{12}^{\ell-1}}  \,, \label{eq:parityeven3pt}
\end{align}
while a basis for the parity-odd structures is given by
  \es{eq:parityodd3pt}{
   r_3 &=  
     \frac{ \<S_3 X_1 X_2 S_3\>^{\ell-1}}
        {X_{12}^{\ell + \frac 12}
        X_{23}^{-\frac{1}{2}} 
        X_{31}^{-\frac{1}{2}}} 
      \left[X_{23} \<S_1 S_3\> \<S_2 X_1 S_3\>
      + X_{13} \<S_2 S_3\> \<S_1 X_2 S_3\> \right] \,, \\
   r_4 &=  
     \frac{ \<S_3 X_1 X_2 S_3\>^{\ell-1}}
        {X_{12}^{\ell + \frac 12}
        X_{23}^{-\frac{1}{2}} 
        X_{31}^{-\frac{1}{2}}} 
      \left[X_{23} \<S_1 S_3\> \<S_2 X_1 S_3\>
      - X_{13} \<S_2 S_3\> \<S_1 X_2 S_3\> \right] \,.
  }
The dependence on $(X_i, S_i)$ in \eqref{FermThreePoint}--\eqref{eq:parityodd3pt} can be derived from the same reasoning as that presented after Eq.~\eqref{ScalarThreePoint}.  The new ingredients here are the transversality of $S_1$ and $S_2$ with respect to $X_1$ and $X_2$, respectively, as well as the Fierz identities
 \es{Fierz}{
  \langle S_1 X_2 S_3 \rangle \langle S_2 X_1 S_3 \rangle &= - \langle S_1 S_2 \rangle \langle S_3 X_1 X_2 S_3 \rangle
   + 2 \langle S_1 S_3 \rangle \langle S_2 S_3 \rangle X_1 \cdot X_2 \,, \\
  \langle S_1  X_3 S_2 \rangle \langle S_3  X_1 X_2 S_3 \rangle &= 
    - 2 \langle S_1 S_3 \rangle \langle S_2  X_1 S_3 \rangle X_2 \cdot X_3 
   - 2 \langle S_2  S_3 \rangle \langle S_1  X_2 S_3 \rangle X_1 \cdot X_3 \,,
 }
which can be used to show that 3-point structures proportional to the left-hand sides of these equalities would be redundant.

When we restrict to the case of identical fermions $\Psi_1 = \Psi_2 = \Psi$, anti-symmetry under $1 \leftrightarrow 2$ places further restrictions on which correlators can be nonvanishing. In particular, if $\ell$ is even then we have the constraint $\lambda_{\psi \psi \cO^-}^4 = 0$, while if $\ell$ is odd then $\lambda_{\psi \psi \cO^+}^1=\lambda_{\psi \psi \cO^+}^2=\lambda_{\psi \psi \cO^-}^3=0$.  (The $\pm$ here only serves as a reminder of the parity of the operators $\cO$ that contribute to each structure in \eqref{FermThreePoint}.)  In other words, even-spin operators have two structures of even parity and one structure of odd parity, while odd-spin operators have a single parity-odd structure. Further, for any spin the Grassmann nature of fermions requires $\<\Psi(X_1,S_1) \Psi(X_2,S_2) \cO_\ell(X_3,S_3)\>^* = - \<\Psi(X_1,S_1) \Psi(X_2,S_2) \cO_\ell(X_3,S_3)\>$, implying that all 3-point coefficients $\lambda_{\psi_1 \psi_2 \cO}^a$ must be pure imaginary.

Let us now consider the general structure of 4-point functions. Scalar 4-point functions are constrained by conformal symmetry to have the form
 \es{ScalarFour}{
\<\Phi_1(X_1)\Phi_2(X_2)\Phi_3(X_3)\Phi_4(X_4)\> = \left(\frac{X_{14}}{X_{13}}\right)^{\frac{\Delta_3-\Delta_4}{2}} \left(\frac{X_{24}}{X_{14}}\right)^{\frac{\Delta_1-\Delta_2}{2}} \frac{g(u,v)}{X_{12}^{\frac{\Delta_1+\Delta_2}{2}} X_{34}^{\frac{\Delta_3+\Delta_4}{2}}},
 }
where $g(u,v)$ is an arbitrary function of the conformal cross ratios $u = \frac{X_{12} X_{34}}{X_{13} X_{24}}$ and $v = \frac{X_{23} X_{14}}{X_{13} X_{24}}$.

Similarly, fermion 4-point functions must take the general form
 \es{FermionFourPoint}{
    \<\Psi_1(X_1,S_1)\Psi_2(X_2,S_2)&\Psi_3(X_3,S_3)\Psi_4(X_4,S_4)\> 
       \\
       &= \left(\frac{X_{14}}{X_{13}}\right)^{\frac{\Delta_3-\Delta_4}{2}} 
         \left(\frac{X_{24}}{X_{14}}\right)^{\frac{\Delta_1-\Delta_2}{2}} 
         \frac{ \sum_I t_I g^I(u,v)}{X_{12}^{\frac{\Delta_1+\Delta_2+1}{2}} X_{34}^{\frac{\Delta_3+\Delta_4+1}{2}}} \,,
  }
where the $t_I$ are a basis of tensor structures that can appear in the 4-point function. There are many choices of bases, but one convenient way to organize the structures is in terms of their properties under various exchanges of the coordinates. 

In general, we find that there are 8 independent structures of even parity that may appear. Anticipating applications to the bootstrap, we will choose 4 structures to be symmetric under the exchange $1\leftrightarrow 3$, and 4 to be anti-symmetric. A basis for the symmetric structures is:
\be
t_1 &=& \frac{\<S_1 S_3\> \<S_2 [X_1,  X_3] S_4\> }{2 X_1 \cdot X_3} + \frac{\<S_2 S_4\> \<S_1 [X_2,  X_4] S_3\> }{2 X_2 \cdot X_4 } \,, \nn\\
t_2 &=& \frac{\<S_1 X_2 S_3\> \<S_2 X_1 S_4\>}{X_1 \cdot X_2} - \frac{\<S_1 X_4 S_3\> \<S_2 X_1 S_4\>}{X_1 \cdot X_4} - \frac{\<S_1 X_2 S_3\> \<S_2 X_3 S_4\>}{X_2 \cdot X_3} + \frac{\<S_1 X_4 S_3\> \<S_2 X_3 S_4\>}{X_3 \cdot X_4} \,, \nn\\
t_3 &=& \frac{\<S_1 X_2 S_3\> \<S_2 X_1 S_4\>}{X_1 \cdot X_2} + \frac{\<S_1 X_4 S_3\> \<S_2 X_1 S_4\>}{X_1 \cdot X_4} - \frac{\<S_1 X_2 S_3\> \<S_2 X_3 S_4\>}{X_2 \cdot X_3} - \frac{\<S_1 X_4 S_3\> \<S_2 X_3 S_4\>}{X_3 \cdot X_4} \,, \nn\\
t_4 &=& \frac{\<S_1 S_3\> \<S_2 [X_1,  X_3] S_4\> }{2 X_1 \cdot X_3} - \frac{\<S_2 S_4\> \<S_1 [X_2,  X_4] S_3\>}{2 X_2 \cdot X_4}, \nn\\
\ee
and a basis for the anti-symmetric structures is:
\be
t_5 &=& \<S_1 S_3\>\<S_2 S_4\>, \nn\\
t_6 &=& \frac{\<S_1 [X_2,  X_4] S_3\> \<S_2 [X_1,  X_3] S_4\>}{4 (X_1 \cdot X_3) (X_2 \cdot X_4)} \,, \nn\\
t_7 &=& \frac{\<S_1 X_2 S_3\> \<S_2 X_1 S_4\>}{X_1 \cdot X_2} + \frac{\<S_1 X_4 S_3\> \<S_2 X_1 S_4\>}{X_1 \cdot X_4} + \frac{\<S_1 X_2 S_3\> \<S_2 X_3 S_4\>}{X_2 \cdot X_3} + \frac{\<S_1 X_4 S_3\> \<S_2 X_3 S_4\>}{X_3 \cdot X_4} \,, \nn\\
t_8 &=& \frac{\<S_1 X_2 S_3\> \<S_2 X_1 S_4\>}{X_1 \cdot X_2} - \frac{\<S_1 X_4 S_3\> \<S_2 X_1 S_4\>}{X_1 \cdot X_4} + \frac{\<S_1 X_2 S_3\> \<S_2 X_3 S_4\>}{X_2 \cdot X_3} - \frac{\<S_1 X_4 S_3\> \<S_2 X_3 S_4\>}{X_3 \cdot X_4} \,. \nn\\
\ee

When we restrict to the case that all fermions are identical, $\Psi_1 = \Psi_2 = \Psi_3 = \Psi_4 = \Psi$, there are additional constraints on the allowed structures coming from exchange symmetries. Some of these are highly nontrivial and lead to the bootstrap conditions discussed in the next section. However, there are also trivial constraints on the allowed tensor structures coming from exchanges that leave the cross-ratios $u$ and $v$ invariant: $\{1,2\} \leftrightarrow \{3,4\}$, $\{1,3\} \leftrightarrow \{2,4\}$, and $\{1,2\} \leftrightarrow \{4,3\}$. Symmetry under these exchanges then forces 
 \es{gVanishing}{
   g^3 = g^4 = g^8 = 0 \,.
 }
In other words, restricting to identical fermions means that there are only 5 allowed tensor structures.

\subsection{Conformal Blocks}

Now we would like to understand how the 4-point functions described in the previous section can be decomposed into conformal blocks, which sum up the contributions of all descendants of a given primary operator appearing in the $\Psi \times \Psi$ OPE\@.  There are many approaches to computing conformal blocks in $D>2$, including direct summation~\cite{Ferrara:1971vh,Ferrara:1973vz,Ferrara:1973yt,Ferrara:1974nf,Ferrara:1974ny,DO1}, solving the Casimir differential equation~\cite{DO2,DO3,ElShowk:2012ht,Fitzpatrick:2013sya,Hogervorst:2013sma}, pole expansions~\cite{Kos:2013tga,Kos:2014bka}, and evaluating monodromy-projected conformal integrals~\cite{SimmonsDuffin:2012uy,Fitzpatrick:2014oza,Khandker:2014mpa}. We will here adopt the latter formulation, since it will allow us to express the fermion conformal blocks in terms of derivatives of known scalar conformal blocks, similar to the approach of~\cite{Costa:2011dw,Echeverri:2015rwa}.

Let us briefly review the conformal block decomposition of a four-point function of scalars $\<\f_1\f_2\f_3\f_4\>$.
Performing the $s$-channel OPE, one can write the function $g(u,v)$ appearing in the scalar 4-point function \eqref{ScalarFour} as a sum of conformal blocks:\footnote{Note that the correctness of this formula depends, crucially, on the normalization of the function $g$.  In terms of the coordinates $r$ and $\theta$ introduced in \cite{Hogervorst:2013sma}, Eq.~\eqref{gSum} holds provided that $g(u,v) \sim  \frac{(1)_\ell}{(1/2)_\ell} (-1)^\ell (4r)^\Delta P_\ell(\cos \theta)$ as $r \to 0$, with ${\phi_i}$ normalized as in \eqref{TwoPtScalar} and ${\cal O}$ normalized as in \eqref{TwoPointArbitrarySpin} with $c_{\cal O} = 1$. \label{FootnoteNormalization}}
 \es{gSum}{
  g(u,v) = \sum_{\cO} \lambda_{\phi_1 \phi_2 \cO} \lambda_{\phi_3 \phi_4 \cO} g_{\De,\ell;\Delta_{12},\Delta_{34}}(u,v) \,,
 } 
where the sum runs only over primary operators belonging to both the $\phi_1 \times \phi_2$ and $\phi_3 \times \phi_4$ OPEs.  As described in~\cite{SimmonsDuffin:2012uy}, the conformal block of ${\cal O}$ can be obtained from the integral
 \es{eq:scalarblocks}{ 
&\lambda_{\phi_1 \phi_2 O} \lambda_{\phi_3 \phi_4 O} \left(\frac{X_{14}}{X_{13}}\right)^{\frac{\Delta_3-\Delta_4}{2}} \left(\frac{X_{24}}{X_{14}}\right)^{\frac{\Delta_1-\Delta_2}{2}} \frac{g_{\De,\ell;\Delta_{12},\Delta_{34}}(u,v)}{X_{12}^{\frac{\Delta_1+\Delta_2}{2}} X_{34}^{\frac{\Delta_3+\Delta_4}{2}}} 
 \\
 &=\frac{1}{\cN_{\cO}} \int D^3 X_0 \<\Phi_1(X_1) \Phi_2(X_2) \cO_{\ell} (X_0)\> \< \tilde{\cO}_{\ell} (X_0) \Phi_3(X_3) \Phi_4(X_4) \> \big|_{\cM} \,,
}
where $\tilde{\cO}_{\ell}$ is the shadow operator of dimension $3-\Delta$ whose indices are contracted with those of ${\cal O}_\ell$, $\big|_{\cM}$ denotes a monodromy projection, and $\cN_{\cO}$ is a normalization factor.

Similarly, performing the $s$-channel OPE in the fermion 4-point function \eqref{FermionFourPoint}, one can write
 \es{gSumFerm}{
  g^I(u,v) = \sum_{\cO} \sum_{a, b} \lambda_{\psi_1 \psi_2 \cO}^a \lambda_{\psi_3 \psi_4 \cO}^b\, g^{I; ab}_{\De,\ell;\Delta_{12},\Delta_{34}}(u,v) \,,
 }
where the index $I$ runs over 4-point structures, while $a, b$ run over 3-point structures. Similarly to the scalar case, the outer sum in \eqref{gSumFerm} runs over the conformal primaries $\cO$ that belong to both the $\psi_1 \times \psi_2$ and $\psi_3 \times \psi_4$ OPEs.  The inner sum in \eqref{gSumFerm} is new in the fermion case; it is present because, for any ${\cO}$, there are several OPE coefficients that need to be specified, as in \eqref{FermThreePoint}.  In analogy with \eqref{eq:scalarblocks}, the conformal blocks appearing in the fermion 4-point function \eqref{FermionFourPoint} can be expressed as
 \es{eq:fermionblocks}{
        &\sum_{a, b} \lambda_{\psi_1 \psi_2 \cO}^a \lambda_{\psi_3 \psi_4 \cO}^b \left(\frac{X_{14}}{X_{13}}\right)^{\frac{\Delta_3-\Delta_4}{2}} 
           \left(\frac{X_{24}}{X_{14}}\right)^{\frac{\Delta_1-\Delta_2}{2}} 
           \frac{t_I g^{I; ab}_{\De,\ell; \Delta_{12},\Delta_{34}}(u,v)}
              {X_{12}^{\frac{\Delta_1+\Delta_2+1}{2}} X_{34}^{\frac{\Delta_3+\Delta_4+1}{2}}}  \\
        &\qquad\qquad\qquad=\frac{1}{\tilde{\cN}_{\cO}} \int D^3 X_0 
           \<\Psi_1(X_1,S_1) \Psi_2(X_2,S_2) \cO_{\ell} (X_0)\> \\
          &\qquad\qquad\qquad\qquad\qquad\qquad\qquad\times \< \tilde{\cO}_\ell (X_0) \Psi_3(X_3,S_3) \Psi_4(X_4,S_4) \> \big|_{\cM} \,,
  }
where the index $I$ runs over 4-point function tensor structures. Thus, it is clear that if each structure appearing in the 3-point functions $\<\Psi_1 \Psi_2 \cO_{\ell}\>$ in \eqref{eq:parityeven3pt}--\eqref{eq:parityodd3pt} can be written as derivatives of the scalar 3-point functions $\<\Phi_1 \Phi_2 \cO_{\ell}\>$ in \eqref{ScalarThreePoint}, then the fermion conformal blocks can be computed from the known scalar blocks. This is the approach that we take in this paper.

Concretely, the parity-even structures in Eq.~(\ref{eq:parityeven3pt}) can be generated by applying certain linear differential operators to $\<\Phi_1 \Phi_2 \cO_{\ell}\>$.  In constructing these operators, we define 
 \be 
  \frac{\delta }{ \delta X_k} \equiv \Gamma^A \frac{ \partial }{\partial X_{k}^{A}} \,,
 \ee
and note a few useful identities:
 \begin{align}
   \<S_1 \frac{\delta}{\delta X_1} S_2\> \< S_3 X_1 X_2 S_3 \> &= -2 \<S_2 S_3\> \<S_1 X_2 S_3 \>
     \,, \\
   \<S_2 \frac{\delta}{\delta X_2} S_1\> \< S_3 X_1 X_2 S_3 \> &= 2 \<S_1 S_3\> \<S_2 X_1 S_3 \>
     \,, \\
   \<S_1 \frac{\delta}{\delta X_1} \frac{\delta}{\delta X_2} S_2\> \< S_3 X_1 X_2 S_3 \> &= 8 \<S_1 S_3\> \<S_2 S_3 \>  \,, \\
    \<S_1 \frac{\delta}{\delta X_1} S_2\> (X_1 \cdot X_k) &= \<S_1 X_k S_2\>
     = - \<S_2 \frac{\delta}{\delta X_2} S_1\> (X_2 \cdot X_k) \,, \\
    \<S_1 \frac{\delta}{\delta X_1} \frac{\delta}{\delta X_2} S_2\> (X_1 \cdot X_2) &= 5 \<S_1 S_2 \>    \,.
 \end{align}

Note that in order for differential operators on $X_k,S_k$ to be well-defined, they must preserve the ideal generated by the relations $X_k^2=0,X_k S_k=0,\<S_k S_k\>=0$.  This is indeed true for the operators above, though the derivative $\frac{\ptl}{\ptl X_k^A}$ is not well-defined on its own.

It can be checked that the 3-point structures appearing in the $\langle \Psi_1 \Psi_2 \cO_\ell \rangle$ 3-point function function can be written in terms of the structure appearing in the 3-point function of two scalars and a spin-$\ell$ operator.  Explicitly, we have 
 \es{FermFromScalarThree}{
  \frac{r_a}{X_{12}^{\frac{\Delta_1 + \Delta_2 - \Delta - \ell+1}{2}} 
     X_{23}^{\frac{\Delta_2 - \Delta_1 + \Delta + \ell}{2}} 
     X_{31}^{\frac{\Delta_1 - \Delta_2 + \Delta + \ell}{2}}} = \cD_a 
    \left[\frac{\<S_3 X_1 X_2 S_3\>^{\ell}}{X_{12}^{\frac{\Delta_1 + \Delta_2 - \Delta + \ell}{2}} 
            X_{23}^{\frac{\Delta_2 - \Delta_1 + \Delta + \ell}{2}} 
            X_{31}^{\frac{\Delta_1 - \Delta_2 + \Delta + \ell}{2}}} \right] \,,
 }
where we defined the differential operators
 \es{DDefs}{
    \cD_1 &\equiv \<S_1 S_2\> \Pi_{\frac12,\frac12} \,, \\
    \cD_2 &\equiv -\frac{1}{4\ell(\De-1)} \< S_1 \frac{\delta}{\delta X_1} \frac{\delta}{\delta X_2} S_2\> 
       \Pi_{-\frac12,-\frac12} + \frac{(\De+\De_1 +\De_2-\ell-4)(\De-\De_1-\De_2-\ell+1)}{4\ell(\De-1)} \cD_1 \,, \\
   \cD_3 &\equiv 
      \frac{1}{2(\De-1)}   \left[ \<S_1 \frac{\delta}{\delta X_1} S_2\> \Pi_{-\frac12,\frac12} 
          - \<S_2 \frac{\delta}{\delta X_2} S_1\> \Pi_{\frac12,-\frac12}  \right] \,,\\
    \cD_4 &\equiv \frac{1}{2\ell }  \left[ \<S_1 \frac{\delta}{\delta X_1} S_2\> \Pi_{-\frac12,\frac12} 
          + \<S_2 \frac{\delta}{\delta X_2} S_1\> \Pi_{\frac12,-\frac12}  \right] 
            - \frac{\Delta_1 - \Delta_2}{ \ell } \cD_3 \,,
 }
and $\Pi_{a,b}$ applies a shift to the operator dimensions as $\{\Delta_1,\Delta_2\} \rightarrow \{\Delta_1+a,\Delta_2+b\}$.

Note that $\cD_1$ and $\cD_2$ generate the parity-even 3-point structures, while $\cD_3$ and $\cD_4$ generate the parity-odd ones.  In addition, the operators $\cD_1$, $\cD_2$, and $\cD_3$ are antisymmetric under the exchange $1 \leftrightarrow 2$, while $\cD_4$ is symmetric.  Together with \eqref{ScalarInterchange}, these symmetry properties imply that in the case of identical fermions $\Psi_1 = \Psi_2$, we obtain three-point functions that obey the anti-symmetry requirement in $X_1$ and $X_2$ provided that $\ell$ is even when we use $\cD_1$, $\cD_2$, and $\cD_3$ and that $\ell$ is odd when we use $\cD_4$, in agreement with the discussion following Eq.~\eqref{Fierz}.

Defining 
 \es{DtildeDef}{
    \widetilde{\cD}_a\equiv \cD_a\big|_{1\to 3\,,2\to 4} \,, 
 }
the fermion conformal blocks \eqref{eq:fermionblocks} are given in terms of the known scalar blocks \eqref{eq:scalarblocks} according to the prescription:
 \es{FermFour}{
     &\left(\frac{X_{14}}{X_{13}}\right)^{\frac{\Delta_3-\Delta_4}{2}} 
           \left(\frac{X_{24}}{X_{14}}\right)^{\frac{\Delta_1-\Delta_2}{2}} 
           \frac{t_I g^{I; ab}_{\De,\ell; \Delta_{12},\Delta_{34}}(u,v)}
              {X_{12}^{\frac{\Delta_1+\Delta_2+1}{2}} X_{34}^{\frac{\Delta_3+\Delta_4+1}{2}}} \\
        &\qquad\qquad\qquad\qquad=  \cD_a\widetilde{\cD}_b \left[\left(\frac{X_{14}}{X_{13}}\right)^{\frac{\Delta_3-\Delta_4}{2}} \left(\frac{X_{24}}{X_{14}}\right)^{\frac{\Delta_1-\Delta_2}{2}} \frac{g_{\De,\ell;\Delta_{12},\Delta_{34}}(u,v)}{X_{12}^{\frac{\Delta_1+\Delta_2}{2}} X_{34}^{\frac{\Delta_3+\Delta_4}{2}}} \right] \,.
  }
The explicit formulas for $g^{I; ab}$ in terms of $g$ are rather complicated, and we will not reproduce them here.  Writing the fermionic blocks as derivatives of scalar blocks is useful for numerical applications.  Computing derivatives of scalar blocks is straightforward, for example we use the pole expansion derived in~\cite{Kos:2014bka} to compute the expansion in radial coordinates~\cite{Hogervorst:2013sma} to order $\rho^{60}$.  Derivatives of fermionic blocks are then obtained as a linear transformation on derivatives of scalar blocks.

\section{3D Fermion Bootstrap}
\label{sec:bootstrap}

Let us return to the 4-point function of identical Majorana fermions in a parity preserving 3D CFT\@.  Using \eqref{STos} in  \eqref{FermionFourPoint}, we can write this 4-point function as
\be
\<\psi(x_1,s_1) \psi(x_2,s_2) \psi(x_3,s_3) \psi(x_4,s_4)\> = \frac{1}{x_{12}^{2\De_{\psi}+1} x_{34}^{2\De_{\psi}+1}} \sum_{I} t_I g^I(u,v) \,,
\ee
where $t_I = t_I(x_i,s_i)$ are the 5 different tensor structures that can appear.  Crossing symmetry under $1 \leftrightarrow 3$ gives a constraint
\be
v^{\De_{\psi}+\frac12} \sum_{I} t_I g^I(u,v) = -  u^{\De_{\psi}+\frac12} \sum_{I} t_I\big|_{1\leftrightarrow 3} g^I(v,u) \,,
\ee
where the minus sign on the right-hand side comes from the Grassmann nature of fermions.
In general $t_I\big|_{1\leftrightarrow 3} = M_I^J t_J$ is related by some matrix $M$, but in the previous section we have chosen a basis of 4-point structures such that $t_{I_+}\big|_{1\leftrightarrow 3} = t_{I_+}$ and $t_{I_-}\big|_{1\leftrightarrow 3} = -t_{I_-}$. In this basis the crossing relation becomes
 \es{CrossingInBasis}{
0 &= \sum_{I_+} t_{I_+} \left[v^{\De_{\psi}+\frac12} g^{I_+}(u,v) +  u^{\De_{\psi}+\frac12} g^{I_+}(v,u) \right] \\
 &{}+ \sum_{I_-} t_{I_-} \left[v^{\De_{\psi}+\frac12} g^{I_-}(u,v) -  u^{\De_{\psi}+\frac12} g^{I_-}(v,u) \right] \,,
 }
or, isolating each tensor structure,
 \es{Isolation}{
0 &= v^{\De_{\psi}+\frac12}g^{I_+}(u,v) +  u^{\De_{\psi}+\frac12} g^{I_+}(v,u) \,, \\
0 &= v^{\De_{\psi}+\frac12}g^{I_-}(u,v) - u^{\De_{\psi}+\frac12} g^{I_-}(v,u) \,.
 }

Now, the functions $g^{I_{\pm}}(u,v)$ have a conformal block decomposition:
 \es{guvDecomp}{
   g^{I_{\pm}}(u,v) = 
      \sum_{\substack{\cO^+, \,\ell\,\textrm{even}\\a,b=1,2}} 
      \lambda^a_{\cO^+} \lambda^b_{\cO+} g^{I_{\pm}}_{ab,\De,\ell}(u,v) 
      &+ \sum_{\cO^-, \,\ell\,\textrm{even}} 
      (\lambda^3_{\cO^-})^2 g^{I_{\pm}}_{33,\De,\ell}(u,v) \\
      &\qquad\qquad+ \sum_{\cO^-, \,\ell\,\textrm{odd}} (\lambda^4_{\cO^-})^2 g^{I_{\pm}}_{44,\De,\ell}(u,v) \,, 
 }
where $\cO_{\pm}$ has parity $\pm$, and we have chosen a basis of parity-odd 3-point structures such that the $a=3$ structure only allows even spins and the $a=4$ structure only allows odd spins, as before.  For brevity, we have written $\lambda_{\cO}^a$ instead of $\lambda^a_{\psi\psi\cO}$, and will henceforth continue to do so.  Thus, we can write the crossing equations as
 \es{CrossingRewriting}{
0 =  \sum_{\substack{\cO^+, \,\ell\,\textrm{even}\\a,b=1,2}} \lambda^a_{\cO^+} \lambda^b_{\cO^+} F^{I_{\pm}}_{ab,\De,\ell}(u,v) + \sum_{\cO^-, \,\ell\,\textrm{even}} (\lambda^3_{\cO^-})^2 F^{I_{\pm}}_{33,\De,\ell}(u,v) + \sum_{\cO^-, \,\ell\,\textrm{odd}} (\lambda^4_{\cO^-})^2 F^{I_{\pm}}_{44,\De,\ell}(u,v) \,,
 }
where $F_{ab,\De,\ell}^{I_{\pm}} \equiv v^{\De_{\psi}+\frac12} g_{ab,\De,\ell}^{I_{\pm}}(u,v) \pm u^{\De_{\psi}+\frac12} g_{ab,\De,\ell}^{I_{\pm}}(v,u)$. 

This is the starting point for the usual bootstrap logic. We can exclude assumptions on the spectrum by applying a linear combination of functionals $\alpha_I$:
 \es{crossing}{
0 &=  \sum_{I_{\pm}} \left[\sum_{\substack{\cO^+, \,\ell\,\textrm{even}\\a,b=1,2}} \lambda^a_{\cO^+} \lambda^b_{\cO^+} \alpha_{I_{\pm}} \left(F^{I_{\pm}}_{ab,\De,\ell}(u,v)\right) \right. \\
& \left. + \sum_{\cO^-, \,\ell\,\textrm{even}} (\lambda^3_{\cO^-})^2  \alpha_{I_{\pm}}\left(F^{I_{\pm}}_{33,\De,\ell}(u,v) \right)+ \sum_{\cO^-, \,\ell\,\textrm{odd}} (\lambda^4_{\cO^-})^2  \alpha_{I_{\pm}}\left(F^{I_{\pm}}_{44,\De,\ell}(u,v)\right) \right] \,,
 }
where we look for functionals that satisfy the constraints
\be
-\sum_{a,b=1,2} \lambda^a_{\mathbb{1}} \lambda^b_{\mathbb{1}} \alpha_{I_{\pm}}\left(F^{I_{\pm}}_{ab,0,0}(u,v)\right) &>& 0, \label{properties}\nonumber\\
\alpha_{I_{\pm}} \left(F^{I_{\pm}}_{ab,\De,\ell}(u,v)\right) &\succeq& 0,\qquad \text{for all parity-even operators with $\ell$ even} \nonumber\\
\alpha_{I_{\pm}}\left(F^{I_{\pm}}_{33,\De,\ell}(u,v) \right) &\geq& 0,\qquad \text{for all parity-odd operators with $\ell$ even} \nonumber\\
\alpha_{I_{\pm}}\left(F^{I_{\pm}}_{44,\De,\ell}(u,v) \right) &\geq& 0,\qquad \text{for all parity-odd operators with $\ell$ odd}.\nn\\
\ee
Recall that in our conventions, all $\l_{\cO}^a$ are pure imaginary --- hence the extra sign in the first line above compared to the usual conditions for scalars. The OPE coefficients of the unit operator are given by $\l_{\mathbb{1}}^a = i\de^a_1$.  We search for functionals satisfying these constraints by approximating the search as a semidefinite program and implementing it in the solver \SDPB\ \cite{Simmons-Duffin:2015qma}. Details of this implementation are given in Appendix~\ref{app:sdpb}.

\section{Results}
\label{sec:results}

We can now use the formalism derived in the previous section to derive constraints on the space of CFTs.   In particular, we consider CFTs with a Majorana fermion $\psi$, and focus on scalar operators appearing in the $\psi \times \psi$ OPE\@.  We assume a parity symmetry, so that we can distinguish between parity-odd scalars, which we denote by $\sigma, \sigma', \sigma'', \ldots$ (in increasing order of their dimensions), and parity-even scalars, which we denote by $\epsilon, \epsilon', \epsilon'', \ldots$ (also in increasing order of their dimensions).  

Using the methods described in Section~\ref{sec:bootstrap}, we first derive general bounds on the dimensions of these operators, observing sharp discontinuities that we conjecture to coincide with a 3D CFT containing no relevant scalar operators. We then study the consequences of imposing gaps in the scalar spectrum, making direct contact with the Gross-Neveu models (described below) at large $N$. Finally we study bounds on the coefficient $C_T$ appearing in the two-point function of the canonically-normalized stress tensor.

\subsection{Examples of Fermionic Theories} 
\label{sec:exampleFermion}
 
While presenting our numerical results, it is useful to keep in mind a few simple CFTs that have fermionic operators:
 \begin{itemize}
  \item {\bf Free Theory.}  The theory of a free Majorana fermion $\psi$ has Lagrangian
   \be
    {\cal L} = -\frac 12 \bar \psi \slashed{\partial} \psi \,,
   \ee
where $\bar \psi \equiv \psi^T (i \gamma^0)$ is the conjugate spinor.   The fermionic operator $\psi$ has dimension $\Delta_\psi = 1$.  There are no parity-even scalar operators appearing in the $\psi \times \psi$ OPE\@.  The only parity-odd scalar appearing in $\psi \x \psi$ is $\bar\psi\psi$, which has dimension $2$.  All correlation functions in this theory can be computed via Wick contractions using the free fermion propagator
   \es{freeFermPropag}{
    \langle \psi^\alpha(x_1) \psi_\beta(x_2) \rangle \propto \frac{i (x_{12})^{\a}{}_\b}{|x_{12}|^3} \,. 
   }
   
  \item {\bf Mean Field Theory.} Mean Field Theory is a generalization of the free theory that in general does not have a local Lagrangian description.  Its operators consist of normal-ordered products of a fermionic operator $\psi$ and its derivatives, except that in this case all correlation functions are computed from Wick contractions using the generalized free field propagator
 \es{meanFermPropag}{
    \langle \psi^\alpha(x_1) \psi_\beta(x_2) \rangle \propto \frac{i (x_{12})^{\a}{}_\b}{|x_{12}|^{2 \Delta_\psi + 1}} \,.
  }
  Mean Field Theory is not properly a local QFT because it doesn't have a stress tensor.  However, it satisfies the properties of unitarity and conformal symmetry that we study in this work.
  In the $\psi \times \psi$ OPE there are now parity-even scalar operators with dimensions $2 \Delta_\psi + 1$, $2 \Delta_\psi + 3$, $2 \Delta_\psi + 5$, \ldots, and parity-odd scalars with dimensions $2 \Delta_\psi$, $2 \Delta_\psi + 2$, $2 \Delta_\psi + 4$, \ldots.  In the limit $\Delta_\psi \to 1$, we recover a free theory, plus additional operators proportional to $ \slashed{\partial} \psi$ whose OPE coefficients in the $\psi \x \psi$ OPE vanish.  As in the free theory, the 4-point function of $\psi$ in Mean Field Theory satisfies crossing symmetry.

  \item {\bf Gross-Neveu(-Yukawa) model.}  Another 3D CFT with fermionic operators is the critical point of the Gross-Neveu model~\cite{Gross:1974jv}.  In the Gross-Neveu-Yukawa description, one starts with $N$ Majorana fermions $\psi_i$ (with $i = 1, \ldots, N$ a flavor index) and a parity-odd scalar field $\phi$, with the Lagrangian
   \be
    {\cal L} = -\frac 12 \sum_{i = 1}^N \bar \psi_i (\slashed{\partial} + g \phi ) \psi_i - \frac 12 \partial^\mu \phi\partial_\mu \phi  - \frac 12 m^2 \phi^2 - \lambda \phi^4 \,,
     \label{GNYLag}
   \ee
where $g$ and $\lambda$ are coupling constants. When $N$ is even, this theory can be studied perturbatively in $d = 4 - \epsilon$ dimensions (see, for example~\cite{Moshe:2003xn}).  It has a critical point that can be achieved by appropriately tuning the scalar mass $m^2$ (a fermionic mass term is forbidden by parity symmetry).  This critical point is believed to survive down to $d=3$, where it can also be studied perturbatively in the $1/N$ expansion~\cite{Gracey:1992cp,Derkachov:1993uw,Gracey:1993kc,Moshe:2003xn}. Previous contact between the conformal OPE and the large-$N$ expansion of this model was made in~\cite{Petkou:1996np}.

In the context of this work, we consider a four-point function of the fermionic operator $\psi=\psi_1$. (We leave the study of global symmetries in fermionic CFTs to future work.) The dimensions of operators in this CFT are not currently available at finite $N$.  At large $N$, the dimensions of the lowest few operators are shown in Table~\ref{tab:grossNeveuDimensions} (see also Appendix~\ref{GNYAPPENDIX}).

\begin{table}[!htb]
\centering
\begin{tabular}{c|c|c|l}
  & $\Z_2$ & $O(N)$& $\De$\\
 \hline
$\f$ & $-$ & $\mathbf{1}$ & $1-32/(3\pi^2 N) + \dots$\\
$\psi_i$  & $+$ & $V$ & $1+4/(3\pi^2 N) + \dots$\\
$\bar\psi_{(i}\psi_{j)}$ & $-$ & $\mathrm{Sym}^2(V)$ & $2+32/(3\pi^2 N)+\dots$\\
$\f^2$ & $+$ & $\mathbf{1}$ & $2+32/(3\pi^2 N)+\dots$\\
$\f^3$ & $-$ & $\mathbf{1}$ & $3+64/(\pi^2 N)+\dots$ \\
$\f^k$ & $(-)^k$ & $\mathbf{1}$ & $k+16k(3k-5)/(3\pi^2 N)+\dots$
\end{tabular}
\caption{Representations and one-loop dimensions of low-lying operators in the large-$N$ 3D Gross-Neveu models.  $V$ denotes the vector representation of $O(N)$.  The dimensions of $\psi_i$, and $\f^k$ were computed in \cite{Gracey:1992cp,Derkachov:1993uw,Gracey:1993kc} and reviewed in Appendix~\ref{GNYAPPENDIX}. The dimension of $\bar\psi_{(i}\psi_{j)}$ is computed in Appendix~\ref{GNYAPPENDIX}.}
\label{tab:grossNeveuDimensions}
\end{table}

  \item {\bf The $\cN = 1$ super-Ising model.}  Another example of a 3D CFT with a Majorana fermion $\psi$ is the ${\cal N} = 1$ supersymmetric Ising model.  It is defined as the IR fixed point of the UV Lagrangian
   \be
    {\cal L} = -\frac 12 \bar \psi \slashed{\partial} \psi - \frac 12 \partial^\mu \phi \partial_{\mu}\phi - \frac{g}{2} \phi \bar \psi \psi -\frac{1}{8}\left(g\phi^2 + h \right)^2 \,,
   \ee
(with the parameter $h$ tuned appropriately), which, when setting $m^2 = gh/2$ and $\lambda = g^2/8$, is nothing but the $N=1$ case of the Gross-Neveu-Yukawa model. This Lagrangian has ${\cal N} = 1$ SUSY and can be described in terms of a real superfield $\Sigma = \phi + \bar{\theta} \psi + \frac{1}{2} \bar{\theta}\theta F$ with superpotential
\be
W = h \Sigma + \frac{g}{3} \Sigma^3.
\ee
Note that a superpotential with quadratic or quartic terms in $\Sigma$ is forbidden by parity symmetry, since $\Sigma$ is parity-odd. In the IR this theory is believed to be described by the ${\cal N} = 1$ superconformal algebra $\mathfrak{osp}(1|4)$. 

Just like for the Gross-Neveu-Yukawa CFT, the dimensions of operators in the ${\cal N} =1$ super-Ising model are not known precisely (but may be experimentally probeable \cite{Grover:2013rc}).\footnote{By contrast, the ${\cal N}=2$ super-Ising model has the exactly known dimension $\Delta_{\sigma} = 2/3$ and contact with the 3D $\cN=2$ bootstrap was recently made in~\cite{Bobev:2015vsa,Bobev:2015jxa,Chester:2015qca}.} They must obey, however, relations imposed by supersymmetry such as $\Delta_\psi = \Delta_\sigma + 1/2 = \Delta_{\epsilon} -1/2$ or $\Delta_{\epsilon'} = \Delta_{\sigma'} + 1$, where $\sigma = \phi$, $\sigma' = \phi^3$, $\epsilon = \phi^2$, $\epsilon' = \phi^4$ are the lowest few scalar operators.  The relation $\Delta_{\epsilon} = \Delta_\sigma + 1$ was used in \cite{Bashkirov:2013vya} to derive $\Delta_\sigma \geq 0.565$.  This inequality was obtained by intersecting the supersymmetric line $\Delta_\epsilon = \Delta_\sigma + 1$ with the bootstrap bounds derived from the crossing symmetry of unitary $\Z_2$-invariant CFTs.

 \end{itemize}

\subsection{Universal Dimension Bounds}

Let us start by computing general upper bounds on the dimensions of scalars appearing in the $\psi \x \psi$ OPE\@.  For the moment, we assume only conformal symmetry, parity symmetry, and unitarity.

\label{sec:parityodd}

 \begin{figure}[t!]
    \centering
    Upper bound on lowest parity-odd scalar $\s\in\psi\x\psi$
\includegraphics[width=0.9\textwidth]{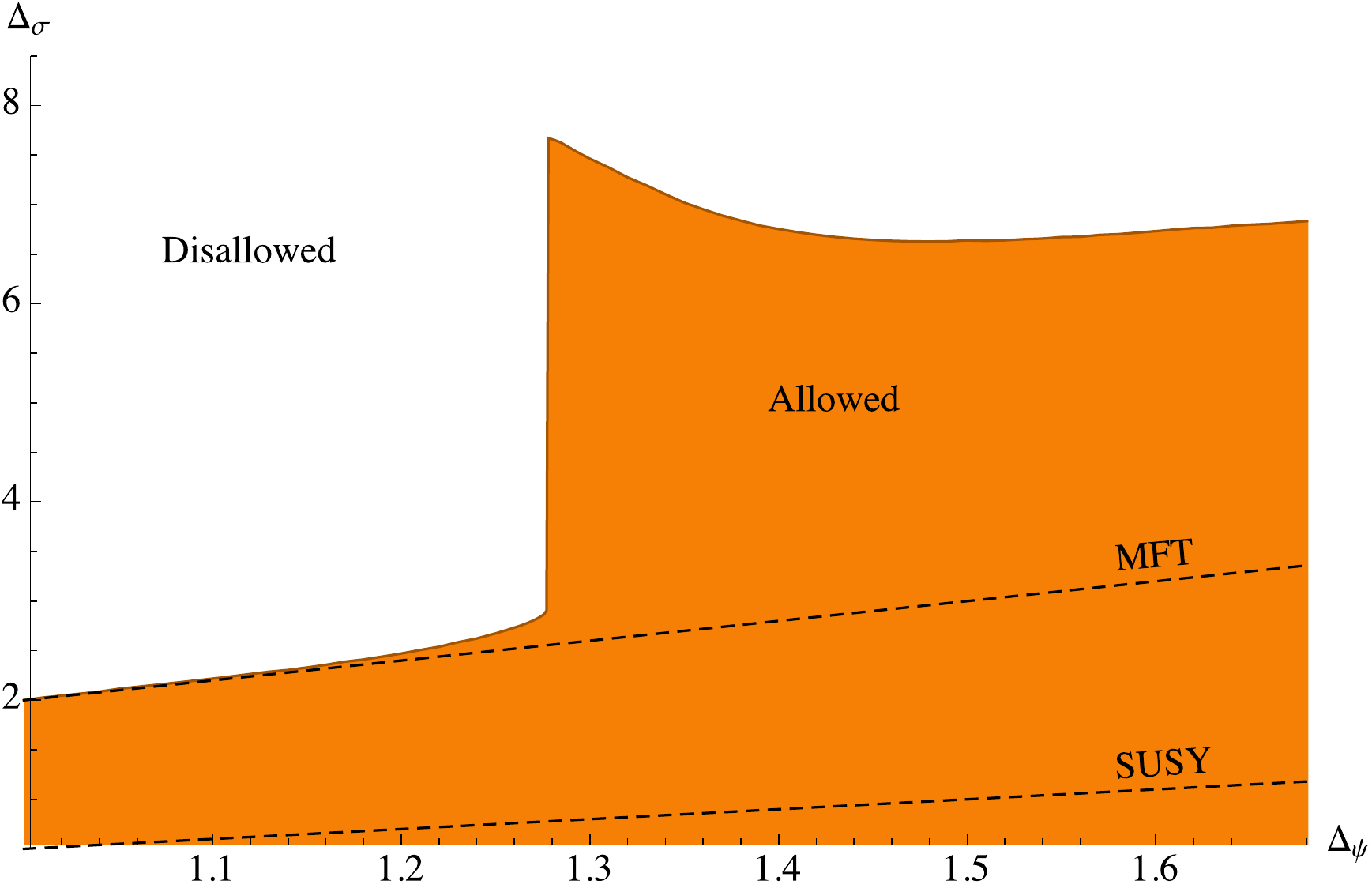}
    \caption{\label{fig:noRelevantParityOdd}
    Upper bounds on the dimension of the lowest dimension parity-odd scalar appearing in the $\psi \times  \psi$ OPE, assuming only conformal symmetry, parity symmetry, and unitarity. The orange region is allowed, and the white region is disallowed. The black dashed line starting at the free theory point $(\Delta_{\psi}, \Delta_\sigma)  = (1, 2)$ gives the relation among dimensions specific to Mean Field Theory, while the dashed line starting at $(\Delta_{\psi}, \Delta_\sigma) = (1, 0.5)$ gives the relation among dimensions expected for $\cN = 1$ SCFTs, assuming $\psi$ is a superdescendant of $\s$.  These bounds are determined using the procedure described in Section~\ref{sec:bootstrap} (see also Appendix~\ref{app:sdpb}) by performing a binary search in $\Delta_{\s}$ with $10^{-3}$ precision. The parameter $\Lambda$ defined in Appendix~\ref{app:sdpb} is given by $\Lambda = 23$. }
  \end{figure}
\subsubsection{The Lowest Dimension Parity Odd Scalar}

In  Figure~\ref{fig:noRelevantParityOdd}, we plot a universal upper bound on $\De_\s$ (the lowest dimension parity-odd scalar) as a function of $\De_\psi$ in any unitary, parity-invariant 3D CFT\@.  The bound starts at the point $(\Delta_\psi, \Delta_\sigma) = (1, 2)$, corresponding to the free theory.  It then grows monotonically with $\Delta_\psi$ up to $\Delta_{\psi} \approx 1.27$, at which point a sharp vertical discontinuity occurs, and the bound jumps  from $\Delta_{\sigma} \approx 2.9$ to $\Delta_\sigma \approx 7.7$.  This striking jump suggests that the value $\De_\psi\approx 1.27$ has special significance. We discuss possible interpretations below.

At the least, we can conclude that any CFT with a fermionic operator of dimension $\Delta_\psi \lesssim 1.27$ must have a relevant parity-odd scalar in the $\psi \times \psi$ OPE\@.  Conversely, a CFT with no relevant parity-odd scalars in the $\psi \times \psi$ OPE must have $\Delta_\psi \gtrsim 1.27$.   In addition, we see that any CFT with a fermion of sufficiently low dimension must have a parity-odd scalar in the $\psi \times \psi$ OPE of dimension smaller than $\approx 7.7$.

\subsubsection{The Lowest Dimension Parity-Even Scalar}
   \begin{figure}[t!]
    \centering
    Upper bound on lowest parity-even scalar $\e\in \psi\x\psi$
\includegraphics[width=0.9\textwidth]{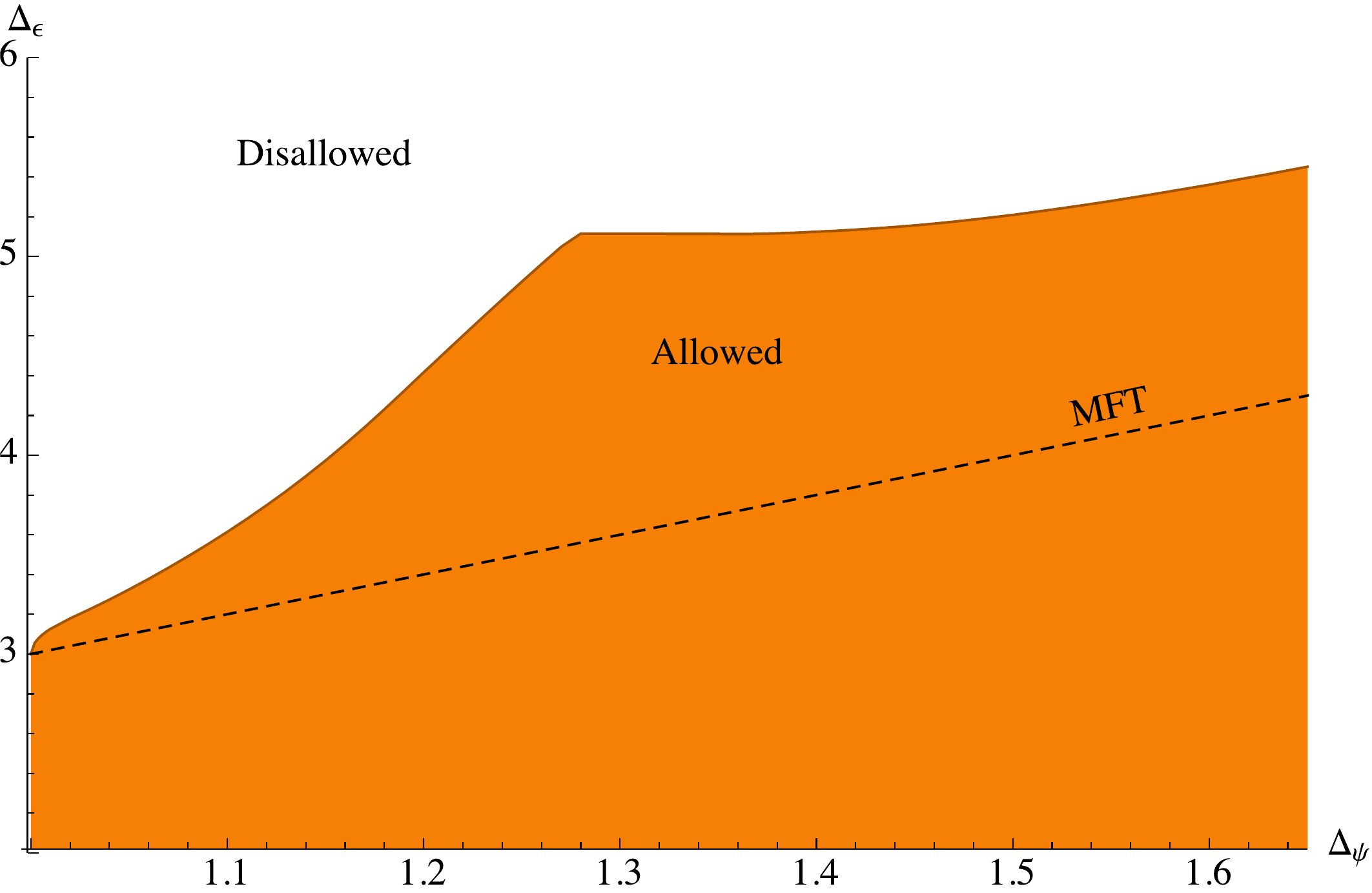}
    \caption{\label{fig:evenScalar}Upper bound on the lowest dimension parity-even scalar appearing in the $\psi \times  \psi$ OPE, as a function of $\Delta_\psi$, assuming only conformal symmetry, parity symmetry, and unitarity. As $\Delta_\psi \to 1$, the bound goes to $\Delta_\epsilon =3$ and has asymptotic behavior $\Delta_\epsilon-3\propto(\Delta_\psi-1)^{1/2}$. The bound has a kink at $\Delta_\psi = 1.27$, which is the same value of $\Delta_\psi$ at which the bound for parity-odd scalars had a discontinuity, see Figure~\ref{fig:noRelevantParityOdd}. This bound was computed with $\Lambda=23$.}
  \end{figure}

In Figure~\ref{fig:evenScalar}, we show an upper bound on $\De_\e$ (the lowest dimension parity-even scalar) in any unitary, parity-invariant 3D CFT\@.  The bound monotonically increases starting from the point $(\Delta_{\psi} , \Delta_\epsilon)= (1, 3)$ up to a value of $\Delta_{\epsilon} \approx 5.1$. At this point, we encounter a change in  slope which occurs at precisely the same value of $\Delta_{\psi}$ as the vertical jump in Figure~\ref{fig:noRelevantParityOdd}.

Note that the free fermion theory does not contain a parity-even scalar of dimension $3$, since the only candidate $\bar\psi \slashed{\partial}\psi$ vanishes by the equations of motion.  However, in Mean Field Theory we have $\Delta_{\epsilon} = 2\Delta_{\psi} + 1$, and hence there exists a continuous family of unitary solutions to crossing symmetry that approach the point $(\De_\psi,\De_\eps)=(1,3)$.  By continuity, our bound cannot move below this point, and indeed it attains this optimal value to high precision.

\subsubsection{A ``Dead End" CFT?}

The kink near $(\De_\psi,\De_\e)\approx(1.27,5.1)$ in Figure~\ref{fig:evenScalar} is reminiscent of the kink in scalar dimension bounds corresponding to the 3D Ising model \cite{ElShowk:2012ht,El-Showk:2014dwa}.  Hence we might guess that there exists a 3D CFT with a fermion of dimension $\De_\psi\approx 1.27$ whose lowest dimension parity-even scalar has dimension $\De_\e\approx 5.1$.

The vertical jump in the parity-odd sector is also reminiscent of a feature previously encountered in scalar dimension bounds. Specifically, Figure~1 of \cite{Kos:2014bka} shows a sharp vertical jump in the bound on $\De_{\s'}$ as a function of $\De_\s$ (assuming that $\De_\e$ saturates its upper bound) in 3d CFTs with a $\Z_2$ symmetry.  That jump went from $\De_{\s'}\approx 2.9$ to $\De_{\s'}\approx 6.8$, and occurred for $0.517\lesssim \De_\s \lesssim 0.52$ (at $\Lambda=11$).  At higher values of $\Lambda$, the $\De_\s$ window shrinks and gives the correct value $\De_\s=0.518151(6)$ in the 3D Ising model. The height of the jump also decreases, e.g.\ to $\De_{\s'}\approx 5.4$ at $\Lambda=19$.  The correct value of $\De_{\s'}$ in the 3D Ising model is approximately $4.5$.

Reasoning by analogy, Figures~\ref{fig:noRelevantParityOdd} and~\ref{fig:evenScalar} lead us to conjecture that there exists a 3D parity-invariant CFT with $\De_\psi \approx 1.27$ and large anomalous dimensions for both the lowest dimension parity-even and parity-odd scalars, perhaps $\De_\e\approx 5.1$, and $3 < \De_\s < 7.7$.  Note that this theory would be a ``dead-end" CFT because it has no relevant scalar operators, giving an example of self-organized criticality~\cite{Bak:1987xua,Bak:1988zz}. In particular, it would be completely attractive under RG flow, and hence would require no tuning to reach criticality (assuming Lorentz-invariance is unbroken).\footnote{Such a theory would also be interesting from the perspective of the AdS/CFT correspondence---an AdS${}_4$ holographic dual of this theory would have no tachyonic scalars and hence all moduli would be fully stabilized. We thank Eva Silverstein for emphasizing this point.} We are not aware of a natural candidate Lagrangian for this theory.\footnote{Several examples of 4D dead-end CFTs were constructed in~\cite{Nakayama:2015bwa}. Their common feature is that they are chiral gauge theories, where mass terms are forbidden by gauge-invariance.} However, the possibility that we have discovered a new ``dead-end" CFT clearly merits further study.

\subsection{Imposing Gaps: The 3D Gross-Neveu Models}

\begin{figure}[t!]
\centering
Allowed $(\De_\psi,\De_\s)$ assuming $\De_{\s'}\geq 2.01,2.03,2.05,2.07,2.09,2.11$
\includegraphics[width=0.9\textwidth]{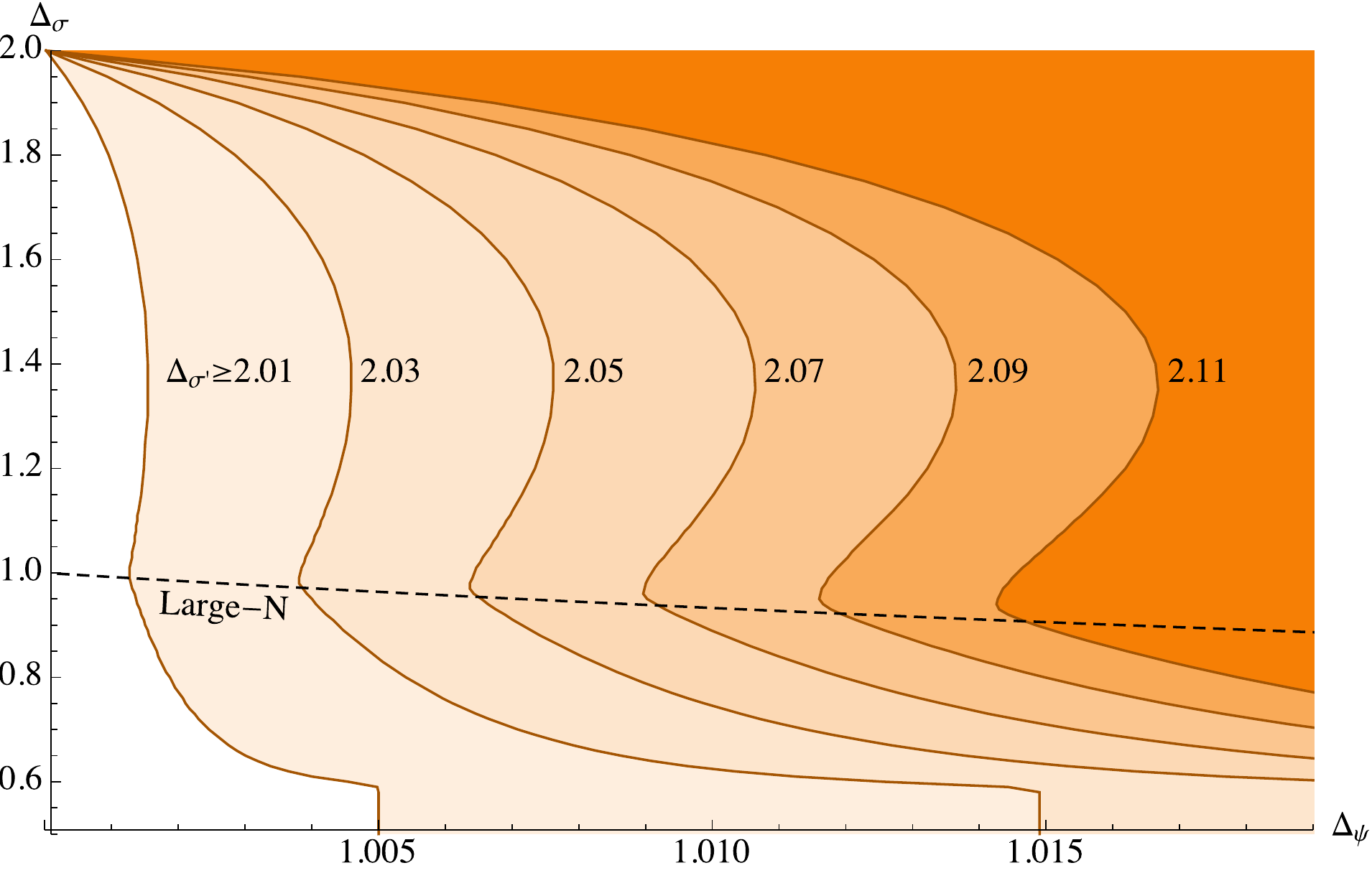}
\caption{
\label{fig:grossNeveuSmallSigPrime}
Allowed values of the dimensions $(\De_\psi,\De_\s)$, assuming $\De_{\s'}\geq \De_{\s'}^\mathrm{min}$ for $\De_{\s'}^\mathrm{min}\in\{2.01,2.03,2.05,2.07,2.09,2.11\}$, computed with $\Lambda=19$.  The regions to the right of their respective curves (shaded orange) are allowed, while the regions to the left are disallowed. The black dashed line shows the relationship between $\De_{\psi}$ and $\De_\s$ using the known 2-loop (for $\De_\s$) and 3-loop (for $\De_\psi$) large-$N$ results in Table~\ref{tab:grossNeveuDimensions} and Appendix~\ref{GNYAPPENDIX}.  The free theory at $(\De_\psi,\De_\s)=(1,2)$ is always allowed.  Below the free theory, there are kinks that closely track the dimensions of operators in the Gross-Neveu models at large $N$. The vertical lines at the bottom of the first two curves ensure consistency of the bounds with Mean Field Theory.
}
\end{figure}

With the most general possible assumptions, we have made contact with the free theory, the limit of Mean Field Theory as $\De_\psi\to 1$, and a conjectured ``dead-end" CFT\@.  Meanwhile, the Gross-Neveu models and $\cN=1$ SUSY Ising model lie well inside the allowed regions in Figures~\ref{fig:noRelevantParityOdd} and~\ref{fig:evenScalar}.  To see them, we must input more information.

A natural choice for the Gross-Neveu models would be to organize operators according to their $O(N)$ representations and use the constraints of $O(N)$ symmetry in the crossing equations, as in \cite{Rattazzi:2010yc,Kos:2013tga,Kos:2015mba}.  We leave this investigation to future work.  For now, we adopt a simpler procedure: we impose gaps in the operator spectrum and study how the bounds change as a function of the gaps.

Specifically, we will use a lower bound $\De_{\s'}\geq \De_{\s'}^\mathrm{min}$ as a proxy for $N$ and try to determine $(\De_\psi,\De_\s)$ as a function of $\De_{\s'}^\mathrm{min}$.  Assuming $\De_{\s'}$ saturates its lower bound, the dependence of $(\De_\psi,\De_\s)$ on $\De_{\s'}$ should be consistent with Table~\ref{tab:grossNeveuDimensions} at large $N$.  Note that because we are considering a single component $\psi=\psi_1$ of the $O(N)$ vector, all representations of $O(N)$ appear in the $\psi\x\psi$ OPE\@.  In particular, we have $\s=\f$ and $\s'=\bar\psi_{(i}\psi_{j)}$ at large $N$.

\begin{figure}[t!]
\centering
Allowed $(\De_\psi,\De_\s)$ assuming $\De_{\s'}\geq 2.1,2.3,2.5,2.7,2.9$
\includegraphics[width=0.9\textwidth]{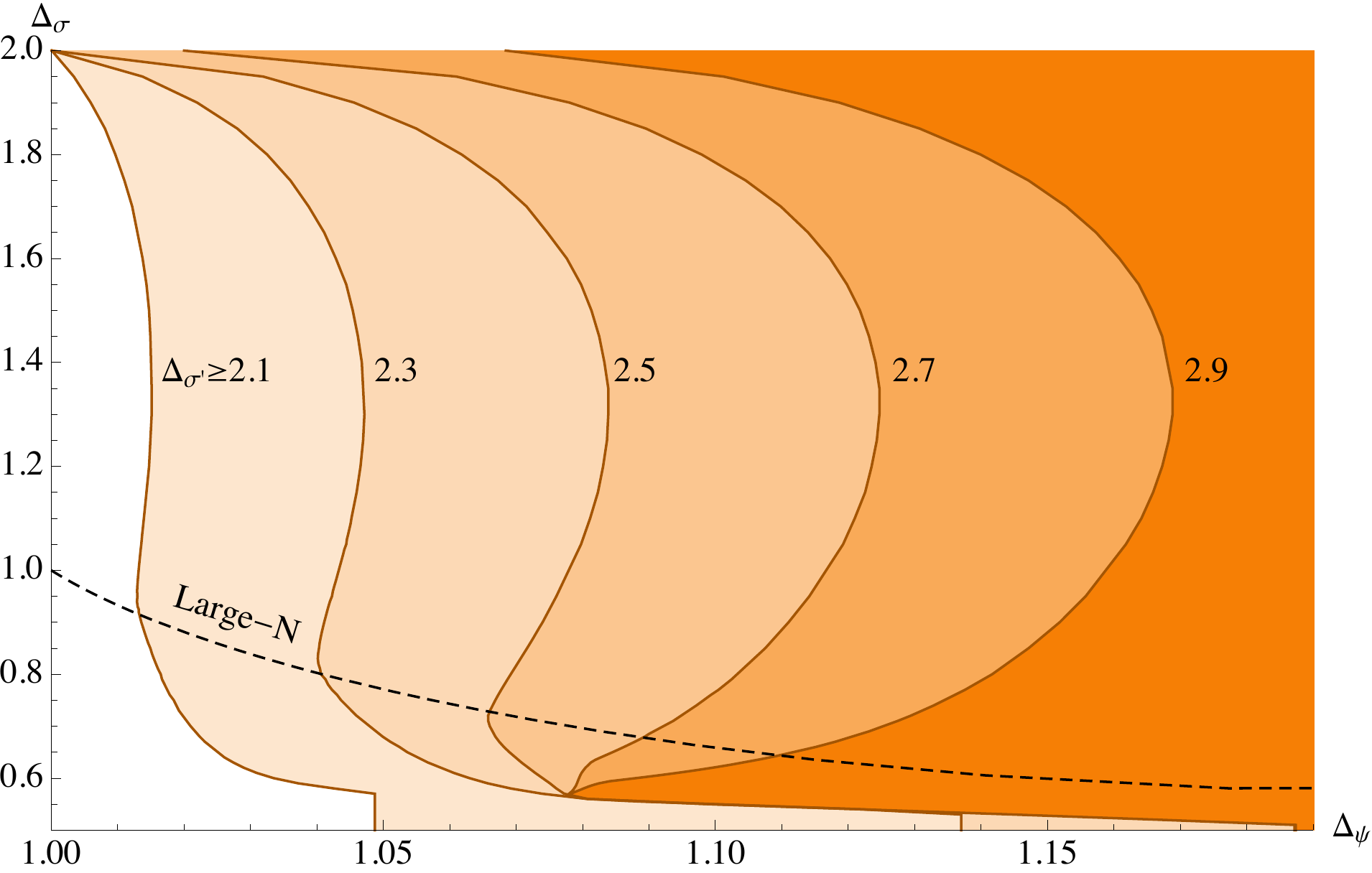}
\caption{
\label{fig:grossNeveuSigPrime2To3}
Allowed values of the dimensions $(\De_\psi,\De_\s)$, assuming $\De_{\s'}\geq \De_{\s'}^\mathrm{min}$ for $\De_{\s'}^\mathrm{min}\in\{2.1,2.3,2.5,2.7,2.9\}$, computed with $\Lambda=19$.  The regions to the right of their respective curves (shaded orange) are allowed, while the regions to the left are disallowed. The black dashed line shows the relationship between $\De_{\psi}$ and $\De_\s$ at 2- and 3-loops at large-$N$.
}
\end{figure}

\begin{figure}[t!]
\centering
Figure~\ref{fig:grossNeveuSmallSigPrime} Kinks vs. 1-loop large-$N$ Gross-Neveu predictions
\includegraphics[width=0.9\textwidth]{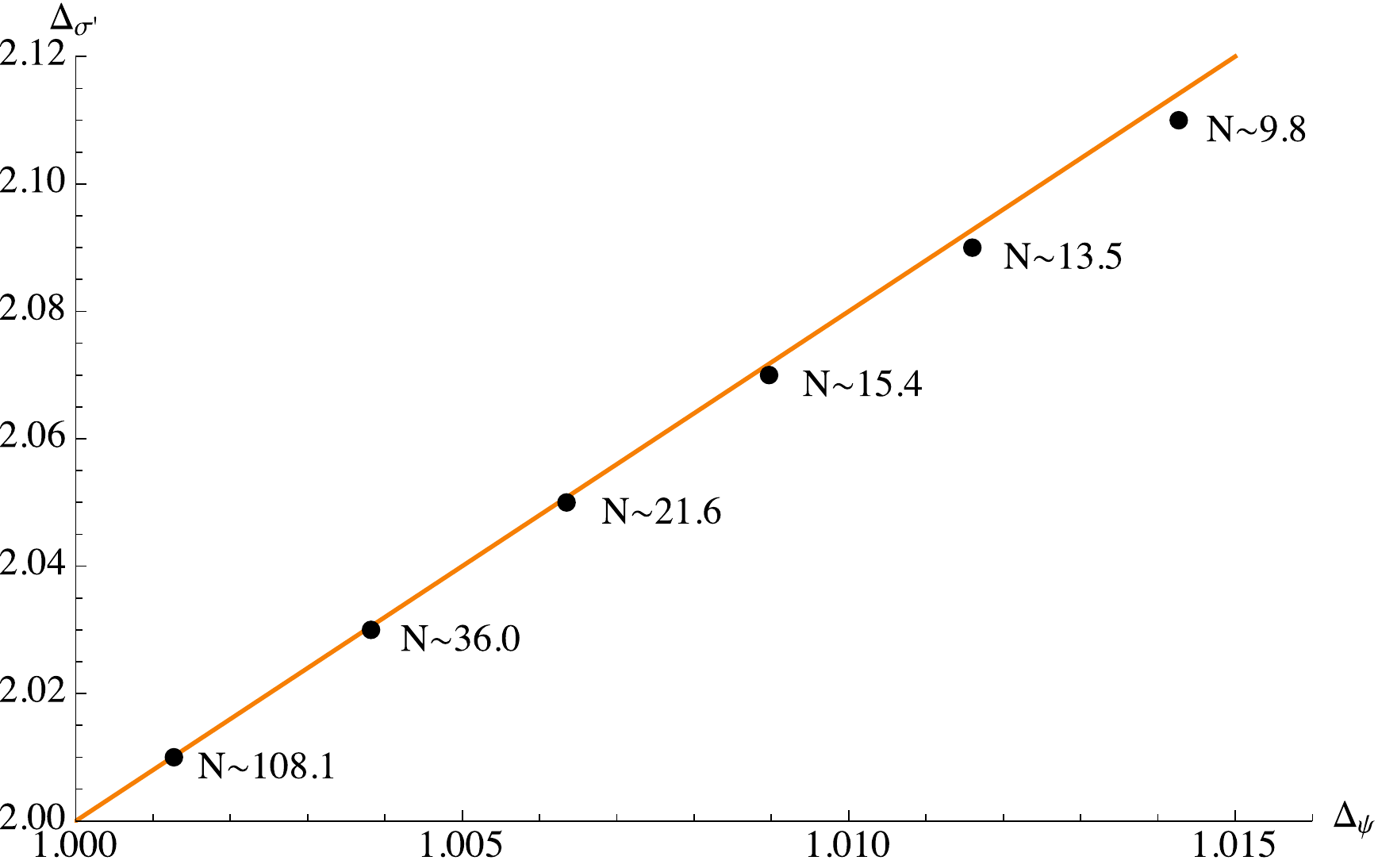}
\caption{
\label{fig:grossNeveuSigmaPrimeVsPsiLargeNComparison}
The positions of the kinks in Figure~\ref{fig:grossNeveuSmallSigPrime} (black points), compared with the 1-loop large-$N$ prediction $\De_{\s'}=8\De_\psi-6$ for the 3D Gross-Neveu models in Table~\ref{tab:grossNeveuDimensions} (orange line).  We also indicate the approximate value of $N$ corresponding to each kink.
}
\end{figure}

In Figures~\ref{fig:grossNeveuSmallSigPrime} and~\ref{fig:grossNeveuSigPrime2To3}, we plot the allowed regions of $(\De_\psi,\De_\s)$ assuming $\De_{\s'}\geq \De_{\s'}^\mathrm{min}$ for several values of $\De_{\s'}^\mathrm{min}$.  All allowed regions are consistent with the free theory at $(\De_\psi,\De_\s)=(1,2)$.  However, the gap in $\De_{\s'}$ has the effect of carving out the allowed region below the free theory, revealing new kinks.  The positions of these kinks closely track the large-$N$ prediction for the Gross-Neveu models, and hence we conjecture that this family of kinks (in the limit $\Lambda\to \oo$) interpolates between the 3D Gross-Neveu models.  In Figure~\ref{fig:grossNeveuSigmaPrimeVsPsiLargeNComparison}, we plot $(\De_{\s'},\De_\psi)$ for the kinks in Figure~\ref{fig:grossNeveuSmallSigPrime}, compared with the large-$N$ prediction at 1-loop, finding excellent agreement.

In Figure~\ref{fig:grossNeveuSigPrime2To3}, we also see that a new kink appears near $(\De_\psi,\De_\s)\approx(1.078,0.565)$ when $\De_{\s'}^\mathrm{min}\gtrsim 2.3$.  This new kink is quite robust to changes in $\De_{\s'}^\mathrm{min}$.  We discuss its possible significance below.

\subsection{Increasing $\De_{\s'}$}

 \begin{figure}[t!]
    \centering
    Allowed $(\De_\psi,\De_\s)$ assuming $\De_{\s'}\geq 3$
\includegraphics[width=0.9\textwidth]{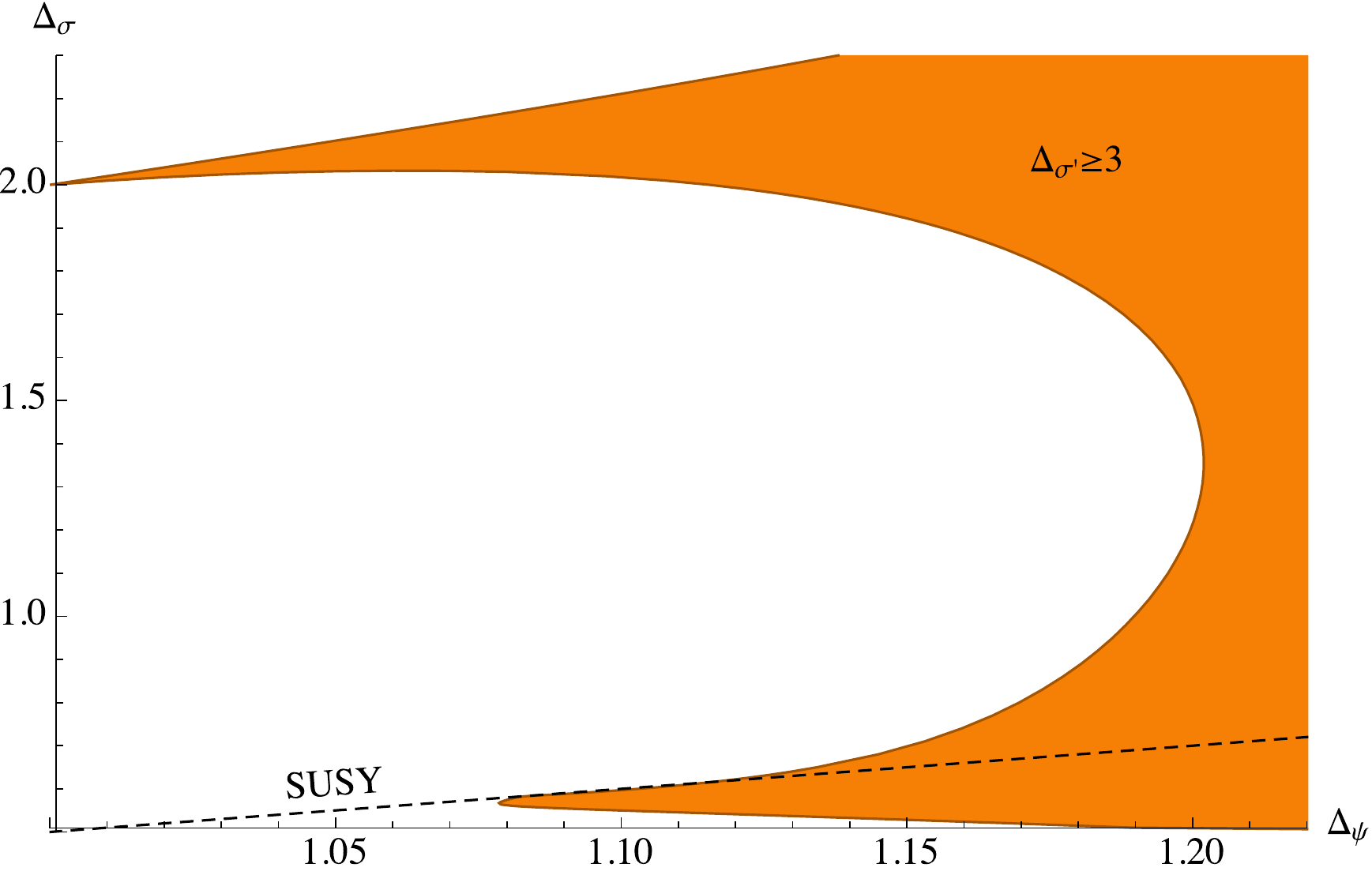}
    \caption{\label{fig:oneRelevant}  Allowed values of $(\De_\psi,\De_\s)$ assuming $\De_{\s'}\geq 3$, computed with $\Lambda=23$.  The orange shaded region is allowed, while the white region is disallowed. The black dashed line shows the SUSY relationship $\De_{\s}=\De_{\psi}-\frac 1 2$.  The kink in the upper-left corner corresponds to the free fermion theory for which $(\Delta_{\psi}, \Delta_\sigma) = (1, 2)$. Figure~\ref{fig:sigmaVsPsiWithSigPrimeGT3Nmax12} zooms in on the second feature, near $(\De_\psi,\De_\s)\approx(1.078,0.565)$.}
  \end{figure}

\begin{figure}[t!]
\centering
Allowed $(\De_\psi,\De_\s)$ assuming $\De_{\s'}\geq 3$ (zoom)
\includegraphics[width=0.9\textwidth]{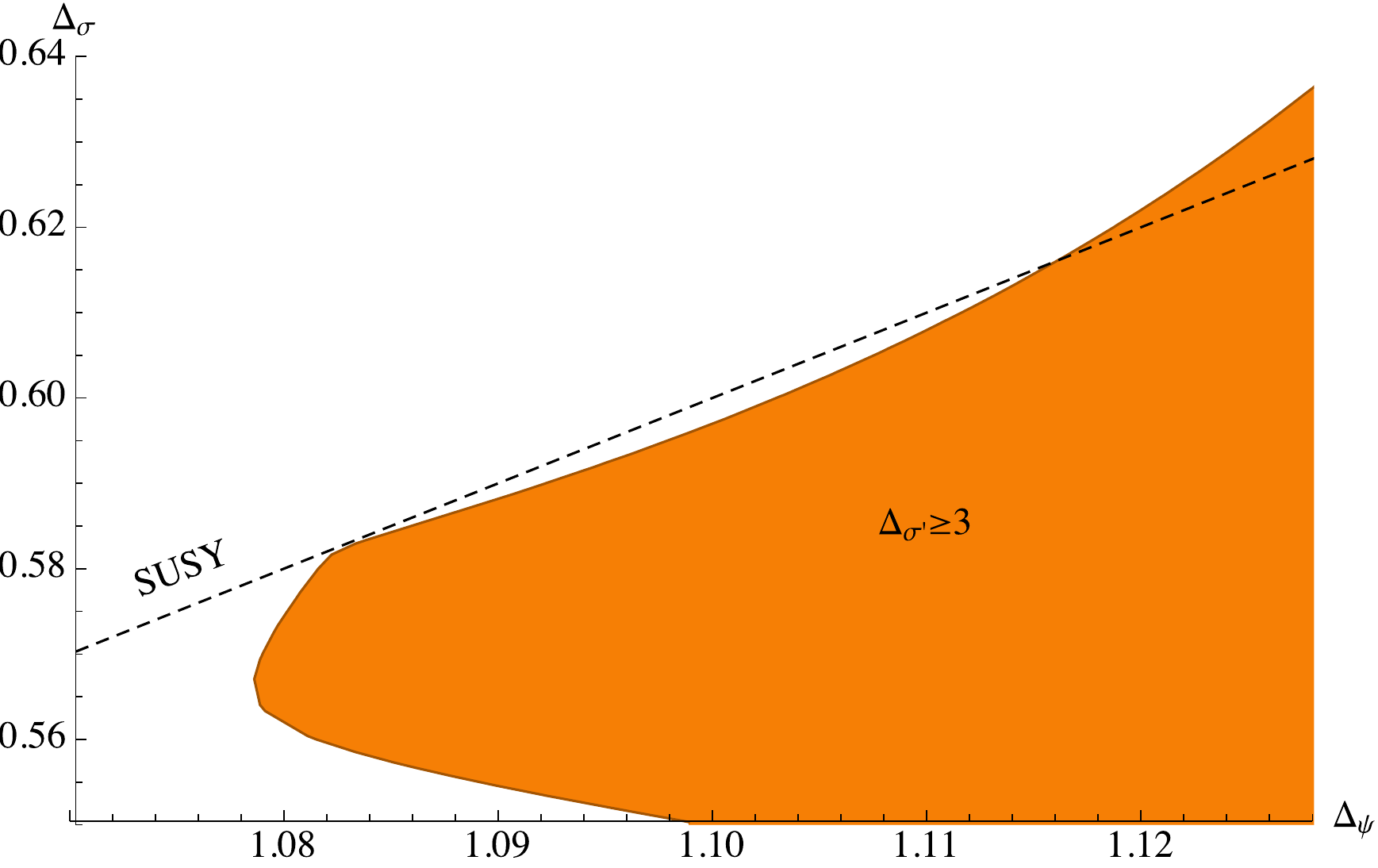}
\caption{
\label{fig:sigmaVsPsiWithSigPrimeGT3Nmax12}
A zoom of Figure~\ref{fig:oneRelevant}: allowed values of $(\De_\psi,\De_\s)$ assuming $\De_{\s'}\geq 3$, computed with $\Lambda=23$.  The orange shaded region is allowed, while the white region is disallowed. The black dashed line shows the SUSY relationship $\De_{\s}=\De_{\psi}-\frac 1 2$. The feature corresponding to the Gross-Neveu models at smaller $\De_{\s'}$ has just crossed the SUSY line near $\De_{\psi}\approx 1.082$.  We observe nothing remarkable when this happens (and the precise dimensions $(\De_\psi,\De_\s,\De_{\s'})$ are sensitive to $\Lambda$). Notice another feature at $(\De_\psi,\De_\s)\approx(1.078,0.565)$. This feature appears already for $\De_{\s'}\ge 2.3$ and persists all the way to $\De_{\s'}\gtrsim 6$ at the same position in $(\De_\psi,\De_\s)$ plane. 
}
\end{figure}

At large $N$, increasing the gap in $\De_{\s'}$ corresponds to decreasing $N$.  We might hope that for big enough $\De_{\s'}$, we could obtain information about the theory with $N=1$, namely the $\cN=1$ supersymmetric Ising model.  In particular, we should identify a feature in the bound that coincides with the line predicted by supersymmetry $\De_\psi=\De_\s+\frac 1 2$.  Unfortunately, we observe nothing particularly special happening along this line.  The kink corresponding to larger $N$ Gross-Neveu models becomes somewhat smooth and crosses the SUSY line when $\De_{\s'}$ is slightly smaller than 3, see the upper kink in Figure~\ref{fig:sigmaVsPsiWithSigPrimeGT3Nmax12}.  If this crossover point corresponded to the $\cN=1$ theory, we would obtain the estimate $\De_\psi \approx 1.082$. However, it is possible that using $\De_{\s'}$ as a proxy for $N$ ceases to work at smaller $N$.  More directly, the operator $\bar\psi_{(i}\psi_{j)}$ (which we have identified with $\s'$) does not actually exist when $N=1$, so it would be unsurprising if the $N>1$ Gross-Neveu kinks are not smoothly connected to the $N=1$ theory using $\De_{\s'}$ as a proxy for $N$.  The precise fate of the Gross-Neveu kinks at small $N$ should become clear when we incorporate the constraints of global symmetry.  We also expect that the $\cN=1$ SUSY Ising model will be easier to isolate using a system of mixed correlators involving both $\psi$ and $\s$.\footnote{An exotic possibility is that the $\cN=1$ SUSY Ising model does not actually exist---that the RG flow from the free $\cN=1$ theory induced by a $\Sigma^3$ superpotential spontaneously breaks SUSY\@.  The theory in the IR of this hypothetical flow would contain a free fermion (by Goldstone's theorem), together with a (possibly empty) interacting sector (perhaps a 3D Ising model).  We thank Juan Maldacena and Igor Klebanov for discussions of this point.}  We leave both of these investigations to future work.

However, our study of increasing $\De_{\s'}$ has revealed an interesting feature in the bound that appears robust:  the kink near $(\De_\psi,\De_\s)\approx(1.078,0.565)$ mentioned in the previous section.  Figure~\ref{fig:oneRelevant} shows the space of allowed dimensions assuming $\De_{\s'}\geq 3$ (equivalently, assuming the theory contains exactly one relevant parity-odd scalar), and this kink appears prominently.  (In fact, it remains present until $\De_{\s'}\gtrsim 6$.) The zoomed-in Figure~\ref{fig:sigmaVsPsiWithSigPrimeGT3Nmax12} makes clear that the lower feature is not consistent with supersymmetry, while the upper feature (mentioned above) is barely incompatible with supersymmetry for $\De_{\s'}\geq 3$.

Thus, we are led to conjecture the existence of a {\it non-supersymmetric\/} 3D parity-invariant CFT with $(\De_\psi,\De_\s)\approx(1.078,0.565)$ and exactly one relevant parity-odd scalar.  The proximity of this theory to the SUSY line may suggest that it is closely related to the $\cN=1$ SUSY Ising model.  For example, suppose the $\cN=1$ SUSY Ising model had a parity-even scalar $\e'$ that was slightly relevant, $\De_{\e'}^{\cN=1}\lesssim 3$.  If $\e'$ is not the top component of a scalar supermultiplet, then deforming the theory by this operator, flowing to the IR, and tuning masses appropriately would yield a non-supersymmetric fixed-point with dimensions very close to those of the SUSY theory.  This possibility could be tested with the bootstrap by identifying the $\cN=1$ SUSY Ising theory, determining the dimension and OPE coefficients of $\e'$, and performing conformal perturbation theory.

\subsection{Central Charge Bounds}
   \begin{figure}[t!]
    \centering
    Lower bound on $C_T$
\includegraphics[width=0.9\textwidth]{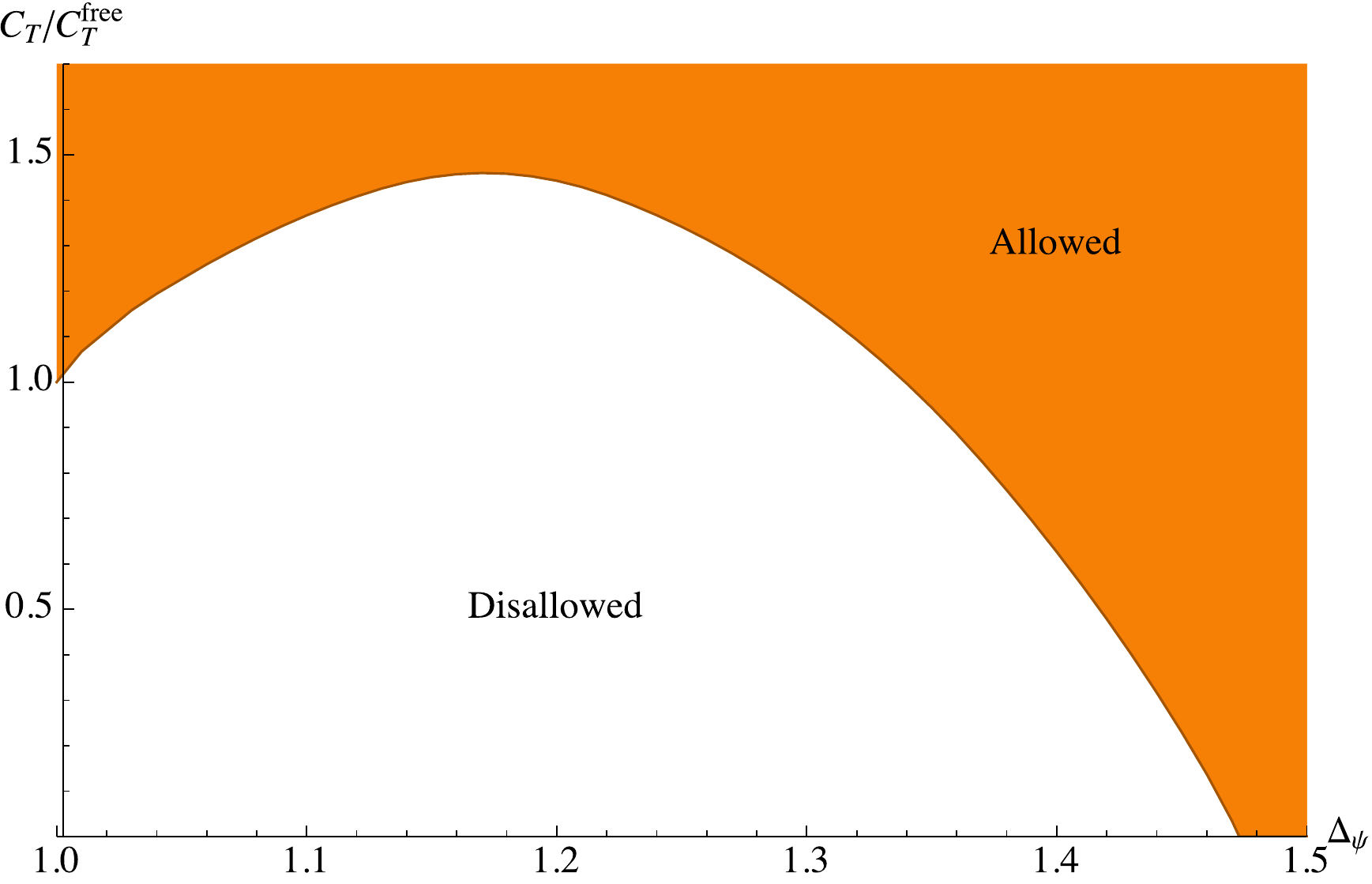}
    \caption{\label{fig:centralCharge}Lower bound on the central charge $C_T$ of a unitary CFT containing a fermion with dimension $\Delta_\psi$. As $\Delta_\psi \to 1$, the bound approaches the free theory value of $C_T$. The bound goes to zero at $\Delta_\psi =1.47$. For larger values of $\Delta_\psi$ the SDP is infeasible and therefore does not produce any bound. This bound was computed with $\Lambda = 23$. }
  \end{figure}
  
Having explored bounds on operator dimensions, we finally turn to the ``central charge" $C_T$ which appears in OPE coefficients of the stress tensor conformal block.  We will place a general lower bound on $C_T$ as a function of $\De_\psi$.

The two-point correlation function of the stress tensor is fixed by conformal invariance to take the form \eqref{Spin2Coords} (with $\Delta = 3)$ up to an overall coefficient.  Let us write 
  \begin{equation}
\label{eq:stresstensor2pt}
  \langle T_\text{can}^{\mu \nu}(x_1) T_\text{can}^{\rho \sigma} (x_2)\rangle = \dfrac{C_T}{(4\pi)^2} \dfrac{1}{x_{12}^6} \left[ \frac 12 \left( I^{\mu\rho}(x_{12}) I^{\nu\sigma}(x_{12}) + I^{\mu\sigma}(x_{12}) I^{\nu\rho}(x_{12})\right) - \frac 13 \eta^{\mu\nu} \eta^{\rho\sigma}   \right]
  \end{equation}  
where $T_\mathrm{can}^{\mu\nu}$ denotes the canonically normalized stress tensor, which participates in the Ward identity for translations as follows:
 \es{StressWard}{
  \frac{\partial}{\partial x^\mu}  \langle T_\text{can}^{\mu\nu}(x) \cO_1(x_1) \ldots \cO_n(x_n) \rangle 
    + \sum_{i=1}^n \delta(x - x_i) \frac{\partial}{\partial x_i^\nu}
   \langle \cO_1(x_1) \ldots \cO_n(x_n) \rangle = 0.
 }
The free boson and free Majorana fermion have $C_T^\text{free} = 3/2$.

The Ward identity (\ref{StressWard}) determines the OPE coefficients $\lambda^a_{T, \text{can}}$ in the 3-point functions in Eq.~\eqref{FermThreePoint}.  As we show in Appendix~\ref{WARD},  we have 
 \es{GotlambdaMainText}{
   \lambda_{ T, \text{can}}^1 = \frac{3 i (\Delta_\psi - 1)}{8 \pi} \,, \qquad
    \lambda_{T, \text{can}}^2 = -\frac{3 i}{4 \pi}  \,.
 }
In our setup, however, our normalization of operators appearing in the $\psi \times \psi$ OPE depends only on $\Delta$ and $\ell$ and is otherwise independent of the details of the CFT we study.  In particular, we can use the normalization of the blocks explained in Footnote~\ref{FootnoteNormalization}, which is equivalent to requiring that the 2-point function of the stress tensor (parity-even operator with $\Delta=3$ and $\ell=2$)  is normalized as in \eqref{TwoPointArbitrarySpin} or \eqref{Spin2Coords} with $c_{\cO} = 1$.  Comparing \eqref{Spin2Coords} with \eqref{eq:stresstensor2pt}, we find 
 \es{TRelation}{
  T^{\mu\nu} = \frac{2 \pi}{\sqrt{C_T}} T^{\mu\nu}_\text{can} \,,
 }
and consequently
 \es{RescalelamT}{
   \lambda_{T}^1 =\lambda_{T, \text{can}}^1 \frac{2\pi}{\sqrt{C_T}}\,,\qquad
   \lambda_{T}^2 = \lambda_{T, \text{can}}^2 \frac{2\pi}{\sqrt{C_T}}\,.
 }
 
We can put a lower bound on $C_T$ as follows. In the sum rule \eqref{crossing}, we isolate the contribution of the parity-even spin-2 operator with $\Delta =3$:
\begin{align}
\lambda^a_{T} \lambda^b_{T} F^{I_{\pm}}_{ab,3, 2}(u,v) = &-\sum_{a,b=1,2}\l_{\mathbb{1}}^a\l_{\mathbb{1}}^bF_{ab,0,0}^{I_\pm}(u,v)   - \sum_{\substack{\cO^+, \,\ell\,\textrm{even}\\a,b=1,2}} \lambda^a_{\cO^+} \lambda^b_{\cO^+} F^{I_{\pm}}_{ab,\De,\ell}(u,v) \notag\\
&- \sum_{\cO^-, \,\ell\,\textrm{even}} (\lambda^3_{\cO^-})^2 F^{I_{\pm}}_{33,\De,\ell}(u,v) - \sum_{\cO^-, \,\ell\,\textrm{odd}} (\lambda^4_{\cO^-})^2 F^{I_{\pm}}_{44,\De,\ell}(u,v), \label{eq:crossingct}
\end{align}
where the summation over parity-even operators now excludes the stress energy tensor and the identity operator, whose contributions we wrote separately. We now search for a functional $\alpha$ such that:
\be
\label{eq:constraintsonfunctionalforcentralcharge}
-\sum_{a,b=1,2} \lambda^a_{T, \text{can}} \lambda^b_{T, \text{can}} \alpha_{I_{\pm}}\left(F^{I_{\pm}}_{ab,3,2}(u,v)\right) &=& 1, \nonumber\\
\alpha_{I_{\pm}} \left(F^{I_{\pm}}_{ab,\De,\ell}(u,v)\right) &\succeq& 0,\qquad  \text{$\forall \Delta \ge \Delta_\ell$, $\ell$ even}\,, \nonumber\\
\alpha_{I_{\pm}}\left(F^{I_{\pm}}_{33,\De,\ell}(u,v) \right) &\geq& 0,\qquad \text{$\forall \Delta \ge \Delta_\ell$, $\ell$ even}\,, \nonumber\\
\alpha_{I_{\pm}}\left(F^{I_{\pm}}_{44,\De,\ell}(u,v) \right) &\geq& 0,\qquad \text{$\forall \Delta \ge \Delta_\ell$, $\ell$ odd}.
\ee
Here, $\De_\ell$ is the lower bound on the dimension of a spin-$\ell$ operator, set by unitarity. Eqs.~\eqref{eq:crossingct} and \eqref{RescalelamT} then imply:
\be
\frac{(2\pi)^2}{C_T} \le - \alpha_{I_\pm}[F_{11,0,0}^{I_\pm}(u,v)]\,,
\ee
where we have used $\l^a_{\mathbb{1}}=i\de^a_1$.
Finding a functional $\alpha$ obeying \eqref{eq:constraintsonfunctionalforcentralcharge} places a lower bound on $C_T$. To make the bound as strong as possible, we search for an $\alpha$ satisfying the relations \eqref{eq:constraintsonfunctionalforcentralcharge} that minimizes $-\alpha_{I_\pm}[F_{11,0,0}^{I_\pm}(u,v)]$. This is slightly different from our procedure for setting bounds on dimensions, where it was enough just to find a functional satisfying certain constraints. Nevertheless, the additional task of finding a functional whose action on a given vector is minimal can again be efficiently performed using \SDPB.

Our central charge lower bound as a function of $\Delta_\psi$ is shown in Figure~\ref{fig:centralCharge}. We normalize $C_T$ by dividing by  its value in the free fermion theory, $C_T^{\text{free}} = 3/2$. The bound has similar features to analogous bounds on $C_T$ coming from scalar 4-point functions in four dimensional CFTs \cite{Poland:2010wg,Rattazzi:2010gj,Poland:2011ey}. As the fermion dimension approaches its free theory value, $\Delta_\psi \to 1$, the bound on $C_T$ also approaches its free theory value. For larger values of $\Delta_\psi$ the bound becomes stronger, reaching a maximum. In this case, the position of the maximum does not coincide with the features observed in the bounds for $\Delta_\sigma$ and $\Delta_\epsilon$, and does not seem to play an important role as it did in the studies of 3D Ising model. At even greater values of $\Delta_\psi$ the bound goes to zero. After that point, the SDP problem described by \eqref{eq:constraintsonfunctionalforcentralcharge} is infeasible, i.e.~it is not possible to find an $\alpha$ satisfying the constraints in \eqref{eq:constraintsonfunctionalforcentralcharge}. Thus, we obtain no bound on $C_T$ for those values of $\Delta_\psi$, beyond the obvious $C_T\ge 0$.

\section{Discussion}
\label{sec:discussion}

In this work, we set up the 3D fermion bootstrap and explored its numerical implications. We first developed an embedding space formalism suitable for describing fermionic correlators. We found that conformal blocks for identical spin-1/2 operators are given by the action of certain differential operators on conformal blocks for scalars. Using these operators, together with differential operators that relate integer spin correlators to scalar blocks \cite{Costa:2011dw}, one can further determine the conformal blocks for any 4-point function in 3D.

On the numerical side, we have foremost shown that the bootstrap can extract rigorous constraints from four-point functions of non-scalar operators. We have obtained general bounds on dimensions of low-lying operators in 3D CFTs with fermions and a small number of relevant scalars. We also obtained general bounds on the central charge $C_T$.  Our results not only provide rigorous constraints on the operator spectrum of CFTs with fermionic operators, but also show numerous features reminiscent of those found when applying the bootstrap to four-point functions of scalar operators.

One interesting feature revealed by the fermionic bootstrap is the kink in the parity-even bound in Figure~\ref{fig:evenScalar}, coinciding with the apparent decoupling of the leading parity-odd scalar in Figure~\ref{fig:noRelevantParityOdd}. We do not yet know the correct interpretation of this feature, but it is intriguing that it may point to the existence of a 3D fermionic CFT with no relevant scalar operators.  If such a theory is responsible for the kink in Figure~\ref{fig:evenScalar}, it would contain a primary spinor operator $\psi$ of dimension $\Delta_\psi \approx 1.27$, and the lowest parity-even scalar appearing in the $\psi \times \psi$ OPE would have dimension $\Delta_\epsilon \approx 5.1$. The possibility that these features reveal a ``dead-end" 3D CFT that gives an example of self-organized criticality merits further study.

Other interesting features occur when we impose a gap to the second relevant parity-odd scalar. By varying its dimension between 2 and 3, we observe a sequence of kinks in the $(\Delta_{\psi}, \Delta_{\sigma})$ plane shown in Figures~\ref{fig:grossNeveuSmallSigPrime} and~\ref{fig:grossNeveuSigPrime2To3}. When the gap is very close to 2, their locations match beautifully onto the dimensions in the Gross-Neveu models at large $N$, seen clearly in Figure~\ref{fig:grossNeveuSigmaPrimeVsPsiLargeNComparison}. At larger values of the gap, we expect that the kink locations make precise predictions in small-$N$ Gross-Neveu models. We also observe the appearance of a new discontinuity in the allowed region at $(\De_\psi,\De_\s)\approx(1.078,0.565)$, which is robust against making the second parity-odd scalar irrelevant.

 In order to better understand if these discontinuities correspond to specific CFTs or SCFTs, one could pursue three immediate steps:\footnote{Given that the same steps would have confirmed that the 3D Ising model populates a ``corner'' in the allowed space of dimensions, even without knowing any critical exponents a \textit{priori}, we can be hopeful that the same will happen for the $\cN =1$ super-Ising model. } 
 \begin{itemize}
 \item One could hope to extend the relation between fermionic conformal blocks and scalar conformal blocks to fractional dimensions. By numerically studying the fermionic crossing-equations in different dimensions, one could compare the evolution of the discontinuities to the results from a perturbative $\epsilon$-expansion (similar to~\cite{El-Showk:2013nia,Bobev:2015jxa}).   
 \item It is straightforward to extend our analysis to constrain fermionic theories with an $O(N)$ global symmetry. As $N$ is varied, we can track the evolution of the bounds on operators in each $O(N)$ representation and again compare with results from the large-$N$ expansion for the Gross-Neveu models. Such a comparison could help confirm that the kinks in Figure~\ref{fig:grossNeveuSigPrime2To3} correspond to the fixed-points of the Gross-Neveu model with a small number of flavors and in particular determine which kinks in our family correspond to integer values of $N$. 

 \item Finally, in order to better understand whether theories live at these discontinuities it would be fruitful to extend our analysis to mixed four-point functions containing both a fermionic operator $\psi$ and a scalar operator $\phi$. This will allow us to impose gaps in the fermionic spectrum, opening up the possibility to obtain isolated islands in the space of operator dimensions, as was seen for scalar correlators in~\cite{Kos:2014bka,Kos:2015mba}. We anticipate that this analysis will be particularly useful for isolating the $\cN=1$ super-Ising model. E.g., these mixed correlators would allow us to determine the fermionic spectrum in the OPE $\phi\times\psi$, enabling us to probe the existence of a conserved supercurrent in the spectrum.

 \end{itemize}
 
We hope to report on these further investigations in future work.

\section*{Acknowledgements}
We thank Chris Beem, Shai Chester, Sheer El-Showk, Simone Giombi, Igor Klebanov, Daliang Li, Juan Maldacena, David Meltzer, Miguel Paulos, Leonardo Rastelli, Slava Rychkov, David Shih,  Eva Silverstein, Andy Stergiou, Balt van Rees, and Alessandro Vichi for discussions. This work was supported by the US NSF under grant No.~PHY-1418069 (LI, SSP, and RY), DOE grant number DE-SC0009988 (DSD), and  NSF grant PHY-1350180 (FK and DP). DSD is supported in part by a William D. Loughlin Membership at the Institute for Advanced Study. In addition, DP, SSP, and DSD acknowledge the support of NSF Grant No. PHY-1066293 and the hospitality of the Aspen Center for Physics. We also thank the organizers and participants of the Back to the Bootstrap workshops at the University of Porto (2014) and the Weizmann Institute (2015). The computations in this paper were run on the Feynman and Della clusters supported by Princeton University, the Omega and Grace computing clusters supported by the facilities and staff of the Yale University Faculty of Arts and Sciences High Performance Computing Center, as well as the Hyperion computing cluster supported by the School of Natural Sciences Computing Staff at the Institute for Advanced Study.

\appendix

\section{Group Theory for 3D Spinors}
\label{app:grouptheory}
The 3D Lorentz group $\SO(2,1)$ has a double cover which is $\SL(2,\bR) \simeq \Sp(2,\bR) \simeq \SU(1,1)$. For us the $\Sp(2,\bR)$ formulation is convenient. It is clear that the smallest irreducible representation is a fundamental of $\Sp(2,\bR)$, which has two real components. This describes a Majorana fermion in 2+1 dimensions.

The Lorentz algebra is
\be
[M^{\mu\nu}, M^{\rho\sigma}] = i (\eta^{\mu \rho} M^{\nu\sigma} + \eta^{\nu \sigma} M^{\mu \rho} - \eta^{\mu \sigma} M^{\nu \rho} - \eta^{\nu \rho} M^{\mu \sigma} )
\ee
In the case of $\SO(2,1)$, we take the signature to be $\eta^{\mu\nu} = \textrm{diag}(-1,1,1)$ and we have 3 generators $J = \frac12 \epsilon_{ab} M^{ab} = M^{12}$, $K_a= M^0{}_a$, where $a,b \in \{1,2\}$.

In the fundamental representation these generators can be written as
\be
J = \left( \begin{array}{ccc} 0&0&0\\0&0&-i\\0&i&0 \end{array} \right), \qquad K_1 = \left( \begin{array}{ccc} 0&0&-i\\0&0&0\\-i&0&0 \end{array} \right), \qquad K_2 = \left( \begin{array}{ccc} 0&i&0\\i&0&0\\0&0&0 \end{array} \right) \,,
\ee
which satisfy the algebra
\be
[J,K_1] = i K_2, \qquad [J,K_2]= - i K_1, \qquad [K_1,K_2]=-i J \,,
\ee
and preserve the metric $\eta M + M^T \eta = 0$. Here $J$ performs a spatial rotation and $K_a$ perform boosts. As usual, the rotation generators are Hermitian while the boost generators are anti-Hermitian.

The fundamental generators of $\Sp(2,\bR)$ (acting on $\psi^\alpha$) satisfy the same algebra and can be written as:
\be
J = \frac12 \left( \begin{array}{cc} 0&i\\-i&0 \end{array} \right), \qquad K_1 = \frac12 \left( \begin{array}{cc} -i&0\\0&i \end{array} \right), \qquad K_2 = \frac12 \left( \begin{array}{cc} 0&i\\i&0 \end{array} \right),
\ee
which preserve a symplectic tensor $\Omega M^{\mu\nu} + (M^{\mu\nu})^T \Omega = 0$, where $\Omega_{\alpha\beta} = \Omega^{\alpha\beta}= \left( \begin{array}{cc} 0&1\\-1&0 \end{array} \right)$. 

The (equivalent) anti-fundamental representation (acting on $\psi_{\alpha} = \Omega_{\alpha\beta}\psi^{\beta}$) transforms with generators $\bar{J} = \Omega J \Omega^{-1} = J$, $\bar{K}_a = \Omega K_a \Omega^{-1} = -K_a$.

The explicit mapping between $\SO(2,1)$ and $\Sp(2,\bR)$ is accomplished via a Clifford algebra:
\be
\gamma^\mu \gamma^\nu + \gamma^\nu \gamma^\mu = 2 \eta^{\mu\nu} \,,
\ee
where we can use the explicit real representation
\be
\gamma^0 = \left( \begin{array}{cc} 0&1\\-1&0 \end{array} \right), \qquad \gamma^1 = \left( \begin{array}{cc} 0&1\\1&0 \end{array} \right), \qquad \gamma^2 = \left( \begin{array}{cc} 1&0\\0&-1 \end{array} \right) \,. \label{gamma}
\ee
The $\Sp(2,\bR)$ fundamental generators are obtained from
\be
({\cal M}^{\mu\nu})^\a{}_\b = -\frac{i}{4} \left(\left[ \gamma^\mu, \gamma^\nu \right]\right)^\a{}_\b \,.
\ee

Note that in our conventions, the index structure on the $\gamma^{\mu}$ matrices defined in \eqref{gamma} is  $(\gamma^{\mu})^{\a}{}_{\b}$. Indices are lowered by multiplying with $\Omega_{\alpha\beta}$ from the left, and raised by multiplying with $\Omega^{\alpha\beta}$ from the right (e.g., $\gamma^{\mu}_{\alpha\beta}\equiv \Omega_{\alpha\gamma}(\gamma^{\mu})^{\gamma}{}_{\beta}$ and $(\gamma^{\mu})^{\alpha\beta} \equiv (\gamma^{\mu})^{\alpha}{}_{\gamma}\Omega^{\gamma\beta}$).

The 3D Lorentzian conformal group $\SO(3,2)$ has a double cover which is $\Sp(4,\bR)$.  We would like to identify which $\Sp(2,\bR)$ subgroup corresponds to the Lorentz rotations described above. We will write the metric as $\eta^{AB} = \textrm{diag}(-1,1,1,1,-1)$, where the first 3 components correspond to $\SO(2,1)$ indices. Then the $\SO(3,2)$ generators $M^{AB}$ which correspond to physical Lorentz rotations and boosts are simply $M^{\mu\nu}$ for $\mu,\nu=0,1,2$.

The mapping between $\SO(3,2)$ and $\Sp(4,\bR)$ is again realized via a Clifford algebra:
\be
\Gamma^A \Gamma^B + \Gamma^B \Gamma^A = 2 \eta^{AB} \,,
\ee
and spinors transform in a representation of $\SO(3,2)$ with generators
\be
(M^{AB})^{I}{}_{J} = -\frac{i}{4} \left[ \Gamma^A, \Gamma^B \right]^{I}{}_{J} \,.
\ee

We can construct a real basis for the $(\Gamma^A)^I{}_J$ matrices explicitly as
\be
\Gamma^0 &=&\left( \begin{array}{cccc} 0&1&0&0\\-1&0&0&0\\0&0&0&-1\\0&0&1&0 \end{array} \right) , \qquad \Gamma^1 = \left( \begin{array}{cccc} 0&1&0&0\\1&0&0&0\\0&0&0&1\\0&0&1&0 \end{array} \right), \qquad \Gamma^2 = \left( \begin{array}{cccc} 1&0&0&0\\0&-1&0&0\\0&0&1&0\\0&0&0&-1 \end{array} \right) \nonumber\\
\Gamma^3 &=& \left( \begin{array}{cccc} 0&0&0&1\\0&0&-1&0\\0&-1&0&0\\1&0&0&0 \end{array} \right), \qquad \Gamma^4 = \left( \begin{array}{cccc} 0&0&0&1\\0&0&-1&0\\0&1&0&0\\-1&0&0&0 \end{array} \right) \,.
\ee
The generators $(M^{AB})^{I}{}_{J}$ in the spinor representation satisfy the $\Sp(4,\bR)$ symplectic constraint $\Omega M^{AB} + (M^{AB})^T \Omega = 0$ with the invariant tensor
\be
\Omega_{IJ} = \Omega^{IJ}=\left( \begin{array}{cccc} 0&0&1&0\\0&0&0&1\\-1&0&0&0\\0&-1&0&0 \end{array} \right) \,.
\ee
Then in this basis the rotation and boost matrices are block diagonal and are given by:
\be
J = \frac12 \left( \begin{array}{cccc} 0&i&0&0\\-i&0&0&0\\0&0&0&i\\0&0&-i&0 \end{array} \right), \quad K_1 = \frac12 \left( \begin{array}{cccc} -i&0&0&0\\0&i&0&0\\0&0&i&0\\0&0&0&-i \end{array} \right), \quad K_2 = \frac12 \left( \begin{array}{cccc} 0&i&0&0\\i&0&0&0\\0&0&0&-i\\0&0&-i&0 \end{array} \right) \,.
\ee
In other words, the upper two components of a $\Sp(4,\bR)$ spinor transform like an $\Sp(2,\bR)$ fundamental, and the lower two components transform like an $\Sp(2,\bR)$ anti-fundamental:
\be
\Psi = \left( \begin{array}{c} \psi^\a \\ \xi_\b \end{array} \right) \,.
\ee

\section{Gross-Neveu-Yukawa model at large $N$}
\label{GNYAPPENDIX}

In this Appendix, we collect known results on the dimensions of low-lying operators in the Gross-Neveu-Yukawa model at its conformal fixed point.  The Lagrangian of the Gross-Neveu-Yukawa model was given in \eqref{GNYLag}.    At the CFT point, one tunes the mass for the scalar field $\phi$ to zero, and one can ignore the quartic scalar interaction as well as the kinetic term for $\phi$.  After rescaling $\phi$, the Lagrangian takes the form 
 \es{LLorentz}{
  {\cal L} = -\frac 12 \sum_{i = 1}^N  \left[ \bar \psi_i \gamma^\mu \partial_\mu \psi_i - i \phi \bar \psi_i \psi_i \right] \,.
 }
Recall that the Majorana condition in Lorentzian signature is $\bar \psi = \psi^T (i \gamma^0)$.  In our conventions, $\gamma^0 = i \sigma_2$, so $\bar \psi = - \psi^T \sigma_2$.

\subsection{Dimensions of $\psi$, $\phi$, and $\phi^2$}

The dimensions of $\psi$, $\phi$, and $\phi^2$ have been computed in \cite{Gracey:1992cp, Derkachov:1993uw,  Gracey:1993kc} at large $N$.  Let us summarize some of these results.  The dimension of the fermion operator $\psi_i$ is known up to order $1/N^3$:
 \es{psiDim}{
  \Delta_\psi = 1 + \frac{4}{3 \pi^2 N} + \frac{896}{27 \pi^4 N^2} 
   + \frac{32 (-668 + 141 \pi^2 + 324 \pi^2 \log 2- 3402 \zeta(3) )}{243 \pi^6 N^3} + O(1/N^4) \,.
 }
The dimension of $\phi$ is 
 \es{DeltaPhi}{
  \Delta_\phi = 1 - \frac{32}{3 \pi^2 N} + \frac{32 (304 - 27 \pi^2) }{27 \pi^4 N^2} + O(1/N^3) \,.
 }
The dimension of $\phi^2$ is
 \es{DeltaPhi2}{
  \Delta_{\phi^2} = 2 + \frac{32}{3 \pi^2 N} - \frac{64 (632 + 27 \pi^2)}{27 \pi^4 N^2} + O(1/N^3) \,.
 }

\subsection{Dimension of $\phi^k$}

To order $1/N$, the dimension of the operator $\phi^k$ can be deduced from the results presented above.  Indeed, at leading order in $1/N$, the dimension of $\phi^k$ equals $k$.  At order $1/N$, there are only two Feynman diagrams contributing, one scaling as $k$ and one as $k(k-1)/2$.  We must therefore have
 \es{Deltaphik}{
  \Delta_{\phi^k} = k + \frac{a k + b k^2}{N} + O(1/N^2) \,,
 }
for some constants $a$ and $b$.  Comparing with \eqref{DeltaPhi}--\eqref{DeltaPhi2}, we have
 \es{GotDeltaphik}{
   \Delta_{\phi^k} = k + \frac{16 k (3k-5)}{3 \pi^2 N} + O(1/N^2) \,.
 }

For $k=3$, for instance, \eqref{GotDeltaphik} gives
 \es{GotDeltaphikParticular}{
  \Delta_{\phi^3} = 3 + \frac{64 }{\pi^2 N} + O(1/N^2) \,.
 }

\subsection{Dimension of $\bar \psi_{(i} \psi_{j)}$}

We are also interested in the dimension of the $O(N)$ symmetric traceless operator $\bar \psi_{(i} \psi_{j)}$, which appears not to have been calculated in the literature.  At leading order in $N$, this operator has dimension $2$.  In the rest of this section, we present the first $1/N$ correction to this result, with the combined answer being
 \es{DeltaSym}{
  \Delta_{\bar \psi_{(i} \psi_{j)}} = 2 + \frac{32}{3 \pi^2} \frac 1N  + O(1/N^2)  \,.
 }

\subsubsection{Setup}

To derive \eqref{DeltaSym}, we find it convenient to work in Euclidean signature.  The Euclidean Lagrangian is the same as \eqref{LLorentz}, with the only difference that we should use Euclidean-signature gamma matrices, which can be taken to be  $\gamma_0 = \sigma_2$, $\gamma_1 = \sigma_1$, $\gamma_2 = \sigma_3$.  The Majorana condition in Euclidean signature is still $\bar \psi = -\psi^T \sigma_2$.

At $N = \infty$, the two-point function of $\psi$ is:
 \es{psiTwoPoint}{
   \langle \psi^\alpha_i(x) \psi^\beta_j(0) \rangle_\infty = \delta_{ij} \frac{i (\gamma_\mu i \sigma_2)^{\alpha \beta} x^\mu}{4 \pi \abs{x}^3} \,.
 }
 In momentum space, this is
  \es{psiTwoPointMom}{
   \langle \psi^\alpha_i(p) \psi^\beta_j(-p) \rangle_\infty = \delta_{ij} \frac{(\gamma_\mu i \sigma_2)^{\alpha \beta} p^\mu}{ p^2 } \,.
  }

The effective action for $\phi$ obtained after integrating out the fermions is
 \es{phiEffect}{
  \frac 12 \int d^3x \int d^3 y\, \phi(x) \phi(y) \Pi_\phi (x, y) \,,
 }
with
 \es{GotPiPhi}{
  \Pi_\phi(x, y) =  \frac 14 \sum_{i, j = 1}^N \langle \bar \psi_i \psi_i (x) \bar \psi_j \psi_j(y) \rangle_\infty \,.
 }
Since $\bar \psi \psi = -\psi^T \sigma_2 \psi =i \epsilon_{\alpha \beta} \psi^\alpha \psi^\beta $,
we have
 \es{PiPhi}{
  \Pi_\phi(x, y) = -\frac 14 \epsilon_{\alpha \beta} \epsilon_{\gamma \delta}\langle \psi_\alpha^i \psi_i^\beta(x) \psi_j^\gamma \psi_j^\delta (y) \rangle_\infty
   = -\frac{N}{(4 \pi)^2 \abs{x - y}^4}
 }
In momentum space, 
 \es{PiMom}{
   \Pi_\phi(p) = \frac{N \abs{p}}{16} \,,
 }
because
 \es{FTx4}{
  \int d^3x e^{i p x } \frac{1}{x^4} = - \pi^2 \abs{p} \,.
 }

To leading order in $N$ we can thus use the propagator for $\psi$:   
 \es{Gpsi}{
  G^{\alpha\beta}_{ij}(p) = \langle \psi^\alpha_i(p) \psi^\beta_j(-p) \rangle  =  \delta_{ij} \frac{(\gamma_\mu i \sigma_2)^{\alpha \beta} p^\mu}{ p^2 }  \,.
 }
The propagator for $\phi$ is $D(p) = 1/\Pi(p)$, or 
 \es{Gphi}{
   D(p) = \langle \phi(p) \phi(-p) \rangle = \frac{16}{N \abs{p}}  \,.
 }

\subsubsection{Anomalous Dimension of $\bar \psi_{(i} \psi_{j)}$}

To compute the anomalous dimension of $\bar \psi_{(i} \psi_{j)}$, let us consider the particular case ${\cal O}(x) = i \epsilon_{\alpha \beta} \psi_1^\alpha \psi_2^\beta(x)$.  The dimension of ${\cal O}$ is 
 \es{OScaling}{
  \Delta_{\cal O} = 2 \Delta_\psi + \eta_\text{vertex} \,,
 }  
where, in terms of Feynman diagrams, $\eta_\text{vertex}$ can be extracted as the coefficient of the logarithmic divergence of the vertex correction diagram.  Keeping track of all the numerical factors and using the propagators \eqref{Gpsi} and \eqref{Gphi}, we have
 \es{etaVertex}{
  \eta_\text{vertex} \log \Lambda + \ldots =  \frac 12 \int \frac{d^3 q}{(2 \pi)^3} 
    \frac{ \tr [ \gamma_\mu    \gamma_\nu ] q^\mu q^\nu}{q^4} \frac{16}{N \abs{q}} = \frac{8  \log \Lambda}{\pi^2 N} + \ldots \,,
 }
from which we extract $\eta_\text{vertex} = 8 / (\pi^2 N)$.  Using \eqref{OScaling} and \eqref{psiDim}, we obtain
 \es{dimO}{
  \Delta_{\cal O} = 2 + \left( \frac{8}{3 \pi^2} + \frac{8}{\pi^2} \right) \frac 1N + O(1/N^2) \,,
 }
yielding \eqref{DeltaSym}.

\section{Implementation in \texttt{SDPB}}
\label{app:sdpb}

In this Appendix we provide a description of the numerical implementation of the fermionic bootstrap using \SDPB\ \cite{Simmons-Duffin:2015qma}.  In order to implement a semi-definite program we limit the space of functionals $\alpha_{I_\pm}$ over which we search over in Section~\ref{sec:bootstrap}, to those taking the form,  
\begin{equation}
\alpha_{I_{\pm}}[f] = \sum_{\substack{n \leq m, \\ m+n \leq \Lambda} } a_{mn}^{I_\pm} \partial_z^m \partial_{\overline{z}}^n f(z, \overline{z})
\bigg|_{z=\overline{z} = \dfrac{1}{2}} \,,
\end{equation} 
with $u = z \overline{z}$ and $v =(1-z) (1-\overline{z})$ and have evaluated the function $f$ at the crossing symmetric point $z=\overline{z} = {1}/{2}$.

Applying these functionals to our crossing equation amounts to finding the $(z, \bar z)$ derivatives of functions $g^{I_\pm}$ appearing in the definition of the conformal block \eqref{eq:fermionblocks}. These functions have singularities as $z\to \bar z$, the most divergent of them going as $(z - \bar z)^{-5}$. The singularities come from our choice of basis $\{ t_I\}$; the full conformal block is perfectly regular at $z=\bar z$. To avoid dealing with the divergences, we multiply the crossing equation by $(z - \bar z)^5$ before applying the functional $\alpha$.

In order to determine the derivatives of the conformal blocks $g^{I_\pm}$ for the fermionic four point functions,  we have used a \texttt{Mathematica} script to apply the operators $\cD_a$ to the rational approximation of the scalar conformal blocks presented in \cite{Kos:2013tga}. Thus, the derivatives of the fermionic conformal blocks $g_{\Delta, \ell}^{I_\pm}$ can be written as 
\begin{equation}
\partial_z^m \partial_{\overline{z}}^n \hat{g}_{\Delta, \ell}^{I_\pm}(z, \overline{z})
|_{z=\overline{z} = {1}/{2}}  \approx \chi_\ell(\Delta) p_\ell^{(m, n), I^\pm}(\Delta) \,,
\end{equation}
where $p_{\ell}^{(m, n), I^\pm}(\Delta)$ are polynomials in $\Delta$ and $\chi_\ell(\Delta)$ is a positive function for all values of $\Delta$ above the unitarity bound. The hat in $\hat{g}$ should remind us that we actually multiplied functions $g$ by $(z-\bar z)^5$. Consequently, at the crossing symmetric point we can write
\begin{equation}
\partial_z^m \partial_{\overline{z}}^n \hat{F}_{ab, \Delta, \ell}^{I_\pm}(z, \overline{z})
|_{z=\overline{z} = {1}/{2}}  \approx \chi_{\ell}(\Delta) P_{ab, \ell}^{(m, n), I^\pm}(\Delta) \,,
\end{equation}
where $P_{ab, \ell}^{(m, n), I^\pm}(\Delta)$ for $a, b \in \{1, 2\}$ or $(a, b) = (3, 3)$, $(a, b) =(4,4)$, are linear combinations of the polynomials $p_\ell^{(m, n), I_\pm}$ determined in \texttt{Mathematica} using (\ref{DDefs}) and the rational approximation of the scalar conformal blocks. Using this approximation, we can rewrite (\ref{properties}) and (\ref{eq:constraintsonfunctionalforcentralcharge}) in the form of a polynomial matrix program solvable using \SDPB\ \cite{Simmons-Duffin:2015qma}, 
\be
&&\text{Find } a_{mn}^{I_{\pm}} \text{ such that:}\nn\\
&&-\sum_{a,b=1,2} \lambda_{\cO_0}^a \lambda_{\cO_0}^b Y_{ab,\ell_0}(\De_0) = 1 \,, \nn\\
&&Y_{ab, \ell}(\Delta) \succeq 0\,\,\, \text{ for all parity-even operators with } \ell \text{ even} \,, \nn \\
&&Y_{33, \ell}(\Delta) \geq 0\,\,\, \text{ for all parity-odd operators with } \ell \text{ even} \,, \nn \\
&&Y_{44, \ell}(\Delta) \geq 0\,\,\, \text{ for all parity-odd operators with } \ell \text{ odd} \,, \label{eq:constraints}
\ee
where the $Y_{ab, \ell}$ are polynomials defined as
\be
Y_{ab, \ell} = \sum_{m,n,I_\pm} a_{mn}^{I_{\pm}}  P_{ab, \ell}^{(m, n), I^\pm}
\ee
for $a, b \in \{1, 2\}$ or $(a, b) = (3, 3)$, $(a, b) =(4,4)$. In our applications we take the operator $\cO_0$ on which we normalize to be either the identity operator or the stress-energy tensor. Note that because of the multiplication of crossing equation by $(z - \bar z)^5$, some of the constraints in (\ref{eq:constraints}) are identically zero, or their linear combinations are identically zero, i.e.\ the set of constraints is not linearly independent. This can cause instabilities in \SDPB, making it run indefinitely.  We want to remove such ``flat directions" and give only linearly independent constraints to \SDPB\@. This can be done numerically. We can view the set of constraints (\ref{eq:constraints}) as a matrix with rows labeling the constraints and columns labeling the components of a functional, $a_{mn}^{I_{\pm}}$. We then only need to find the linearly independent rows of the matrix. That can be done for example in \texttt{Mathematica} using the built-in \texttt{RowReduce} function. Notice that this step needs to be done only once for a given $\Lambda$.

The full description of implementing the polynomial matrix program required to find $a_{mn}^{I_{\pm}}$ can be found in the \SDPB\ manual \cite{Simmons-Duffin:2015qma}. We have used a \texttt{Mathematica} script to manipulate the fermionic conformal blocks to obtain the matrix input for \SDPB\@. 
 In order to obtain numerically accurate results we have used the parameters presented in Table~\ref{tab:parameters} in our \SDPB\ implementation. For $\Lambda = 19$ generating the input file required by \SDPB\ takes about 30 minutes (on a single core), while solving each semi-definite program takes 25 minutes (allowed points) or 100 minutes (disallowed points) on an 8 core machine. For $\Lambda = 23$ generating the input file required by \SDPB\ takes about 90 minutes while solving each semi-definite program takes 3 hours (allowed points) or 14 hours (disallowed points) on an 8 core machine. 
 \begin{table}[!t]
\centering
\begin{tabular}{|l|c c|}
\hline
$\Lambda$ & 19 & 23  \\
$\numax$ & 20 & 24\\
spins & $S_{19}$ & $S_{23}$ \\
\texttt{precision} & 640 & 960 \\
\texttt{findPrimalFeasible} & True & True \\
\texttt{findDualFeasible} & True & True  \\
\texttt{detectPrimalFeasibleJump} & True & True \\
\texttt{detectDualFeasibleJump} & True & True \\
\texttt{dualityGapThreshold} & $10^{-25}$ & $10^{-40}$\\
\texttt{primalErrorThreshold} & $10^{-25}$ & $10^{-100}$ \\
\texttt{dualErrorThreshold} & $10^{-25}$ & $10^{-40}$ \\
\texttt{initialMatrixScalePrimal} ($\Omega_\cP$) & $10^{20}$ & $10^{40}$ \\
\texttt{initialMatrixScaleDual} ($\Omega_\cD$) & $10^{20}$ & $10^{40}$ \\
\texttt{feasibleCenteringParameter} ($\beta_\mathrm{feasible}$) & 0.1 & 0.1 \\
\texttt{infeasibleCenteringParameter} ($\beta_\mathrm{infeasible}$) & 0.3 & 0.3\\
\texttt{stepLengthReduction} ($\gamma$) & 0.7 & 0.7 \\
\texttt{choleskyStabilizeThreshold} ($\theta$) & $10^{-40}$ & $10^{-40}$  \\
\texttt{maxComplementarity} & $10^{100}$ & $10^{130}$ \\
\hline
\end{tabular}
\caption{Parameters for the computations in this work.  Only \SDPB\ parameters that affect the numerics (as opposed to parameters like \texttt{maxThreads} and \texttt{maxRuntime}) are included.  The sets of spins used are $S_{19}= \{0, 1, 2, \dots, 25\} \cup \{29,30,33,34,37,38,41,42,45,46,49,50\}$ and $S_{23}= \{0, 1, 2, \dots, 25\} \cup \{29,30,33,34,37,38,41,42,45,46,49,50,59,60\}$.}
\label{tab:parameters}
\end{table}

\section{Conformal Ward identities}
\label{WARD}

In this appendix we study the implications of the Ward identity given in Eq. \eqref{StressWard} for correlators containing fermions. One can multiply \eqref{StressWard} by a conformal Killing vector $\xi_\nu$ satisfying $\partial_{(\mu} \xi_{\nu)} \propto \eta_{\mu\nu}$.  In a conformal field theory, the fact that the stress tensor $T^{\mu\nu}$ is symmetric and traceless implies
 \es{MoreGeneralStressWard}{
   \frac{\partial}{\partial x^\mu}  \langle  \xi_{\nu}(x) T^{\mu\nu}(x) \cO_1(x_1) \ldots \cO_n(x_n) \rangle 
    + \sum_{i=1}^n \delta(x - x_i) \xi_{\nu}(x_i) \frac{\partial}{\partial x_i^\nu}
   \langle \cO_1(x_1) \ldots \cO_n(x_n) \rangle = 0  \,.
 }
Taking $x_1 = 0$, ${\cO}_1 = \cO$, and integrating in $x$ over a small enough sphere of radius $\epsilon$ centered at the origin, one can extract the integrated OPE
 \es{OPEInt}{
   \epsilon^2 \int_{S^2} d^2 \hat n \, n_{\mu} \xi_{\nu}(x) T^{\mu\nu}(x)  {\cal O}(0) = i[Q_\xi, {\cal O}](0)\,,
 }
where in deriving the expression we also used Stokes' theorem and $Q_\xi$ is the conserved charge whose associated conserved current is $J^\mu(x) = \xi_\nu(x) T^{\mu\nu}(x)$.  Specializing to Lorentz transformations, translations, special conformal transformations, and dilatations, we simply replace $\xi_{\nu}(x) T^{\mu\nu}(x)$ with
 \es{GenDef}{
  (M_{\nu\rho})_\mu(x) &= x_\rho T_{\mu \nu} - x_\nu T_{\mu \rho} \,, \\
  (P_\nu)_\mu(x) &= - T_{\mu\nu} \,, \\
  (K_\nu)_\mu(x) &= 2 x_\nu x^\rho T_{\mu\rho} - x^2 T_{\mu\nu} \,, \\
  D_\mu(x) &= x^\nu T_{\mu\nu} \,
 } 
in \eqref{OPEInt}, and $Q_\xi$ with $M_{\mu\rho}$, $P_\nu$, $K_\nu$, and $D$, respectively.

We are interested in calculating the OPE coefficient between the stress tensor and a spinor primary field $\psi$.  Using \eqref{psiConf}, Eq.~\eqref{OPEInt} becomes 
 \es{GenInt}{
   \epsilon^2  \int_{S^2} d^2 \hat n\,\hat n^\mu (M_{\nu\rho})_\mu(\epsilon \hat n) \psi(0)&= - \frac{1}{2} \gamma_{\nu\rho} \psi(0) \,, \\
   \epsilon^2 \int_{S^2} d^2 \hat n\,\hat n^\mu (P_{\nu})_\mu(\epsilon \hat n) \psi(0)&= -\partial_\nu \psi(0) \,, \\
   \epsilon^2  \int_{S^2} d^2 \hat n\,\hat n^\mu (K_{\nu})_\mu(\epsilon \hat n) \psi(0)&=0 \,, \\
   \epsilon^2  \int_{S^2} d^2 \hat n\,\hat n^\mu D_\mu(\epsilon \hat n) \psi(0)&= \Delta_\psi \psi(0)\,.
 }

The general form of the OPE $T_{\mu\nu} \times \psi$ is restricted by the tracelessness and conservation of $T_{\mu\nu}$ to take the form
  \es{TpsiOPECan}{
   T_{\mu\nu}  (x) \psi(0)  &=  a \frac{\eta_{\mu\nu}x^2 - 3 x_\mu x_\nu}{\abs{x}^5} \psi(0)
   + b \frac{x_\mu x^\rho \gamma_{\rho \nu} + x_\nu x^\rho \gamma_{\rho \mu}}{\abs{x}^5} \psi(0) + \cdots \,,
 }
for some constants $a$ and $b$.  Using the definitions \eqref{GenDef}, we have
  \es{ChargepsiOPE}{
  x^\mu (M_{\nu\rho})_\mu (x) \psi(0) &= 
    b \frac{x_\rho  x^\sigma \gamma_{\sigma \nu} - x_\nu  x^\sigma \gamma_{\sigma \rho} }{\abs{x}^3} \psi(0) 
    + O(x^0) \,, \\
  x^\mu (P_{\nu})_\mu (x) \psi(0) &=  a \frac{2 x_\nu  }{\abs{x}^3} \psi(0)
   - b \frac{ x^\sigma \gamma_{\sigma \nu} }{\abs{x}^3} \psi(0) 
    + O(x^{-1}) \,, \\
  x^\mu (K_{\nu})_\mu (x) \psi(0) &=  -a \frac{2 x_\nu }{\abs{x}} \psi(0) 
    - b \frac{ x^\sigma \gamma_{\sigma \nu} }{\abs{x}} \psi(0)  
    + O(x) \,, \\
   x^\mu D_\mu (x) \psi(0) &= -a \frac{2}{\abs{x}} \psi(0) +  O(x^0) \,,  
 } 
and so
  \es{IntExplicit}{
   \epsilon^2  \int_{S^2} d^2 \hat n\,\hat n^\mu (M_{\nu\rho})_\mu(\epsilon \hat n) \psi(0)&= - \frac{8 \pi b}{3} \gamma_{\nu\rho} \psi(0) + O(\epsilon) \,, \\
   \epsilon^2 \int_{S^2} d^2 \hat n\,\hat n^\mu (P_{\nu})_\mu(\epsilon \hat n) \psi(0)&= O(\epsilon^0) \,, \\
   \epsilon^2  \int_{S^2} d^2 \hat n\,\hat n^\mu (K_{\nu})_\mu(\epsilon \hat n) \psi(0)&=O(\epsilon^2) \,, \\
   \epsilon^2  \int_{S^2} d^2 \hat n\,\hat n^\mu D_\mu(\epsilon \hat n) \psi(0)&= -8 \pi a \psi(0) + O(\epsilon)\,.
 }
Comparing \eqref{IntExplicit} with \eqref{GenInt}, we identify 
 \es{Gotab}{
   a = -\frac{\Delta_\psi}{8 \pi} \,, \qquad b = \frac{3}{16 \pi} \,.
 }
The final form of the $T\times \psi$ OPE is
   \es{TpsiOPEFinal}{
   T_{\mu\nu}  (x) \psi(0)  &=  -\frac{\Delta_\psi}{8 \pi} \frac{\eta_{\mu\nu}x^2 - 3 x_\mu x_\nu}{\abs{x}^5} \psi(0)
   +  \frac{3}{16 \pi} \frac{x_\mu x^\rho \gamma_{\rho \nu} + x_\nu x^\rho \gamma_{\rho \mu}}{\abs{x}^5} \psi(0) + \cdots \,.
 }

Let us now compare this expression with what we expect from the 3-point function \eqref{FermThreePoint}.  For a parity-even operator, we have
 \es{FermFermOComp}{
  \langle \psi^\b(x_1) \psi^\gamma(x_2) {\cal O}^{\a_1 \ldots \a_{2 \ell}}  (x_3)  \rangle
   &= \lambda_{\cal O}^1 \frac{x_{12}^{\b\gamma}  (x_{31} x_{12} x_{23} )^{(\a_1 \a_2} \cdots (x_{31} x_{12} x_{23} )^{\a_{2\ell-1} \a_{2\ell})} }{\abs{x_{12}}^{2 \Delta_\psi  - \Delta + \ell + 1} \abs{x_{23}}^{\Delta + \ell} \abs{x_{31}}^{\Delta + \ell}} \\
    &+ \lambda_{\cal O}^2 \frac{(x_{13})^{\b (\a_1} (x_{23})^{|\gamma| \a_2} (x_{31} x_{12} x_{23} )^{\a_3 \a_4} \cdots (x_{31} x_{12} x_{23} )^{\a_{2\ell-1} \a_{2 \ell})} }{\abs{x_{12}}^{2 \Delta_\psi  - \Delta + \ell - 1} \abs{x_{23}}^{\Delta + \ell} \abs{x_{31}}^{\Delta + \ell}} \,.
 }
where as usual $x^{\a\b} = x^\mu (\gamma_\mu \Omega)^{\a\b}$.  From the $x_3 \to x_1$ limit of the 3-pt function we can deduce the ${\cal O} \times \psi$ OPE\@.  In this limit, the 3-pt function is 
 \es{FermFermOCompapprox}{
  \langle \psi^\b(x_1) \psi^\gamma(x_2) {\cal O}^{\a_1 \ldots \a_{2 \ell}}  (x_3)  \rangle
   &\approx \lambda_{ \cal O}^1 (-1)^\ell \frac{x_{12}^{\b\gamma}  (x_{31} )^{(\a_1 \a_2} \cdots (x_{31} )^{\a_{2\ell-1} \a_{2\ell})} }{\abs{x_{12}}^{2 \Delta_\psi  + 1}\abs{x_{31}}^{\Delta + \ell}} \\
    &+ \lambda_{\cal O}^2 (-1)^{\ell} \frac{(x_{31})^{\b (\a_1} (x_{21})^{|\gamma| k_2} (x_{31} )^{\a_3 \a_4} \cdots (x_{31})^{\a_{2\ell-1} \a_{2 \ell})} }{\abs{x_{12}}^{2 \Delta_\psi + 1}  \abs{x_{31}}^{\Delta + \ell}} \,.
 }
Using the normalization where $\langle \psi^\a (x) \psi^\b(0) \rangle = i x^{\a\b} / \abs{x}^{2 \Delta_\psi + 1}$, the OPE contribution of $\psi$ then is
 \es{OPEpsi}{
   {\cal O}^{\a_1 \ldots \a_{2 \ell}}  (x_3) \psi^\beta(x_1)  &\sim  i (-1)^{\ell+1}  \lambda_{\cal O}^1\frac{   (x_{31} )^{(\a_1 \a_2} \cdots (x_{31} )^{\a_{2\ell-1} \a_{2\ell})}  }{  \abs{x_{31}}^{\Delta + \ell}  }  \psi^\beta(x_1) \\
     &+i  (-1)^{\ell} \lambda_{ \cal O}^2  \frac{   (x_{31} )^{\beta(\a_1 } \cdots (x_{31} )^{\a_{2\ell-2} \a_{2\ell-1}}  }{  \abs{x_{31}}^{\Delta + \ell}  }  \psi^{\a_{2 \ell})}(x_1) \,,
 }
because this contribution reproduces the 3-pt function in the OPE limit.

Let us now specialize to the case where ${\cal O}_2 = T$ is the canonically normalized stress tensor.  Eq.~\eqref{OPEpsi} is in this limit
 \es{OPEstressFerm}{
     T^{\a_1 \a_2 \a_3 \a_4 }  (x) \psi^\beta(0)  &\sim  
       -i   \lambda_{ T}^1\frac{1}{\abs{x}^5}  x^{(\a_1 \a_2} x^{\a_{3} \a_{4})}    \psi^\beta(0) 
       +i  \lambda_{ T}^2  \frac{1}{\abs{x}^5}  x^{\beta(\a_1 }  x^{\a_{2} \a_{3}}   \psi^{\a_{4})}(0) \,.
 }
Using \eqref{OSptime} we can represent the stress tensor as a rank-2 Lorentz tensor:
 \es{OPEstressFermAgain}{
   T_{\mu\nu}  (x) \psi^\beta(0)  &\sim  \frac 14  (\Omega \gamma_\mu)_{\a_1 \a_2} (\Omega \gamma_\nu )_{\a_3 \a_4} \left[
       -i   \lambda_{T}^1\frac{1}{\abs{x}^5}  x^{(\a_1 \a_2} x^{\a_{3} \a_{4})}    \psi^\beta(0) 
       +i  \lambda_{T}^2  \frac{1}{\abs{x}^5}  x^{\beta(\a_1 }  x^{\a_{2} \a_{3}}   \psi^{\a_{4})}(0) \right] \,.
 } 
For the first term, we can use
 \es{GammaIds}{
  &x^\sigma x^\rho (i \sigma_2 \gamma_\mu)_{\a_1 \a_2} (i \sigma_2 \gamma_\nu )_{\a_3 \a_4}
       (\gamma_\sigma i \sigma_2 )^{(\a_1 \a_2}  (\gamma_\rho  i \sigma_2)^{\a_{3} \a_{4})}
       = \frac 13 x^\sigma x^\rho \left[ \tr (\gamma_\mu \gamma_\sigma) \tr (\gamma_\nu \gamma_\rho)
        + 2 \tr (\gamma_\mu \gamma_\sigma \gamma_\nu \gamma_\rho )\right] \\
      &= \frac 13 \left[4 x_\mu x_\nu - 4 x^2 \eta_{\mu\nu} + 8 x_\mu x_\nu  \right] 
      = -\frac 43 \left( \eta_{\mu\nu} x^2 - 3 x_\mu x_\nu \right)  \,.
 }
For the second term, we have
 \es{SecondGammaId}{
  &x^\sigma x^\rho (i \sigma_2 \gamma_\mu)_{k_1 k_2} (i \sigma_2 \gamma_\nu )_{k_3 k_4} (\gamma_\sigma i \sigma_2)^{i(k_1 }  (\gamma_\rho i \sigma_2)^{k_{2} k_{3}} \psi^{k_4 )} \\
  &=\frac{1}{6} x^\sigma x^\rho \left(2 \gamma_\sigma \gamma_\mu \gamma_\rho \gamma_\nu \psi
   + 2 \gamma_\sigma \gamma_\nu \gamma_\rho \gamma_\mu \psi + \tr (\gamma_\rho \gamma_\mu) \gamma_\sigma \gamma_\nu \psi + \tr (\gamma_\rho \gamma_\nu) \gamma_\sigma \gamma_\mu \psi \right)^i \\
  &= \frac{1}{6} x^\sigma x^\rho \left(- 2 \eta_{\rho \sigma} \gamma_\mu \gamma_\nu \psi 
    + 4 \eta_{\mu\rho} \gamma_\sigma  \gamma_\nu \psi + 2 \eta_{\mu \rho}  \gamma_\sigma \gamma_\nu \psi + \text{$(\mu \leftrightarrow \nu)$} \right)^i \\
  &= \frac{1}{6} \left[ - 4 x^2 \eta_{\mu\nu} \psi + 6 (x_\mu x^\sigma \gamma_{\sigma \nu} + x_\nu x^\sigma \gamma_{\sigma \mu})  \psi + 12 x_\mu x_\nu \psi\right]^i \\
  &=  \left[ - \frac{2}{3} ( x^2 \eta_{\mu\nu} - 3 x_\mu x_\nu) \psi +  (x_\mu x^\sigma \gamma_{\sigma \nu} + x_\nu x^\sigma \gamma_{\sigma \mu})  \psi \right]^i
 }
So:
 \es{TpsiOPE}{
  T_{\mu\nu}  (x) \psi(0)  \sim  \frac{i}6 (2 \lambda_{T}^1 - \lambda_{\psi\psi T}^2)  \frac{\eta_{\mu\nu}x^2 - 3 x_\mu x_\nu}{\abs{x}^5} \psi(0)
   + \frac{i}{4} \lambda_{T}^2 \frac{x_\mu x^\rho \gamma_{\rho \nu} + x_\nu x^\rho \gamma_{\rho \mu}}{\abs{x}^5} \psi(0)  \,.
 }

We can compare \eqref{TpsiOPECan} to \eqref{TpsiOPE} to obtain
 \es{Gotlambda}{
   \lambda_{ T}^1 = \frac{3 i (\Delta_\psi - 1)}{8 \pi} \,, \qquad
    \lambda_{T}^2 = -\frac{3 i}{4 \pi}  \,.
 }

\bibliography{Biblio}{}
\bibliographystyle{utphys}

\end{document}